\documentclass[preprint,12pt]{aastex}
\newcommand{\jhks}{\ensuremath{JHK_s}}
\newcommand{\spitzer}{\textit{Spitzer}}
\newcommand{\ak}{\ensuremath{A_{K_s}}}
\newcommand{\chisq}{\ensuremath{\chi^2}}
\newcommand{\cext}{\ensuremath{C_{ext}(\lambda)}}

\newcommand{\alak}{\ensuremath{A_\lambda/A_{K_s}}}
\newcommand{\paperone}{Paper I}

\begin{document}

\author{Nicholas L.\ Chapman\altaffilmark{1,2},
Lee G.\ Mundy\altaffilmark{1}}

\altaffiltext{1}{Department of Astronomy, University of Maryland, College Park,
MD 20742; chapman@astro.umd.edu}
\altaffiltext{2}{Jet Propulsion Laboratory, California Institute of  Technology,
4800 Oak Grove Drive, MS 301-429, Pasadena, CA 91109}

\title{ Deep $JHK_s$ and \spitzer\ Imaging of Four Isolated Molecular Cloud Cores }

\begin{abstract}

We present observations in eight wavebands from $1.25-24\:\mu$m of four dense
cores: L204C-2, L1152, L1155C-2, and L1228.  Our goals are to study the YSO
population of these cores and to measure the mid-infrared extinction law. With
our combined near-infrared and \spitzer{} photometry, we classify each source in
the cores as, among other things,  background stars, galaxies, or embedded young
stellar objects (YSOs). L1152 contains three YSOs and L1228 has seven, but
neither L204C-2 nor L1155C-2 appear to contain any YSOs.  We estimate an upper
limit of $7\times10^{-5}$ to $5\times10^{-4}$ $L_\odot$ for any undiscovered
YSOs in our cores.  We also compute the line-of-sight extinction law towards
each background star.  These measurements are averaged spatially, to create
\chisq{} maps of the changes in the mid-infrared extinction law throughout our
cores, and also in different ranges of extinction.  From the \chisq{} maps we
identify two small regions in L1152 and L1228 where the outflows in those cores
appear to be destroying the larger dust grains, thus altering the extinction law
in those regions. On average, however, our extinction law is relatively flat
from $3.6$ to $24\:\mu$m for all ranges of extinction and in all four cores. 
From $3.6$ to $8\:\mu$m this law is consistent with a dust model that  includes
larger dust grains than the diffuse interstellar medium, which suggests grain
growth has occurred in our cores.  At $24\:\mu$m, our extinction law is
$2-4\times$ higher than predicted by dust models.  However, it is similar to
other empirical measurements.

\end{abstract}


\section{ Introduction }

In the standard model of low-mass star formation, dense cores are the precursors
of star formation \citep{shu87}.  Therefore, to understand the process of star
formation, one needs to observe the physical conditions of these cores.
Embedded young stellar objects (YSOs) are sometimes detected in these cores.
These YSOs are partially obscured by the dust within the cores.  To accurately
model their physical properties requires knowledge of the quantity of dust and
its composition, since both affect the degree of attenuation, or extinction, of
starlight.  By studying the dust, one can learn about its chemistry and
evolution.  Furthermore, the dust provides a direct measure of the column
density and mass. Molecular tracers can suffer depletion in dense regions,
making mass estimates from them unreliable.

In this paper we use \spitzer{} and deep ground-based \jhks{} observations to
probe the dust properties within four cores: L204C-2, L1152, L1155C-2, and
L1228. Two of our cores, L1152 and L1228, have known protostars while the other
two do not. Our first goal in this paper is to examine the YSO population of
these cores and to look for faint previously unknown protostars in all four
cores. Second, we will compute the extinction law in each core and compare it to
the predictions of three dust models to draw conclusions about the dust
properties within the cores. This paper is organized as follows:
\S\,\ref{sec:observe} describes our observations and data processing pipeline
including source classification, the removal of misidentified background
galaxies, and how we computed line-of-sight (LOS) extinctions to each star. In
\S\,\ref{sec:core-ext} we use these LOS extinctions to create extinction maps
for each core. Then, \S\,\ref{sec:coreysos} explores the YSO content in our
cores. In \S\,\ref{sec:dust} we compute the relative extinction as a function of
wavelength, the extinction law, for each core. We also create maps of the
\chisq{} maps of the difference between the observed and predicted extinction
laws and compare to the extinction maps. Finally, we summarize our results in
\S\,\ref{sec:conclude}.

\section{\label{sec:observe}Observations}

Table \ref{tab:corebasic} lists the location and our assumed  distance to each of
our cores: L204C-2, L1152, L1155C-2, and L1228. These cores were all observed as
part of the \spitzer{} Legacy Science program ``From Molecular Cores to
Planet-forming Disks'' (c2d) \citep{evans03}. Using \spitzer, we re-observed
these cores with $\sim10\times$ the integration time as c2d to look for faint
protostars in the cores. These observations were part of GO program \#3656. We
observed each core in two epochs, where the second epoch was separated from the
first by as little as one day but could be up to almost a month later. L204C-2
is located at an ecliptic latitude where asteroids may be a problem; the
separation of the two observing epochs ensures we can identify these sources.
IRAC ($3.6-8\:\mu$m) and MIPS1 ($24\:\mu$m) were observed separately, resulting
in four \spitzer{} Astronomical Observation Requests (AORs) per core. Our
\spitzer{} AORs are listed in Table \ref{tab:coreobs}.

\begin{figure}
\epsscale{0.9}
\plotone{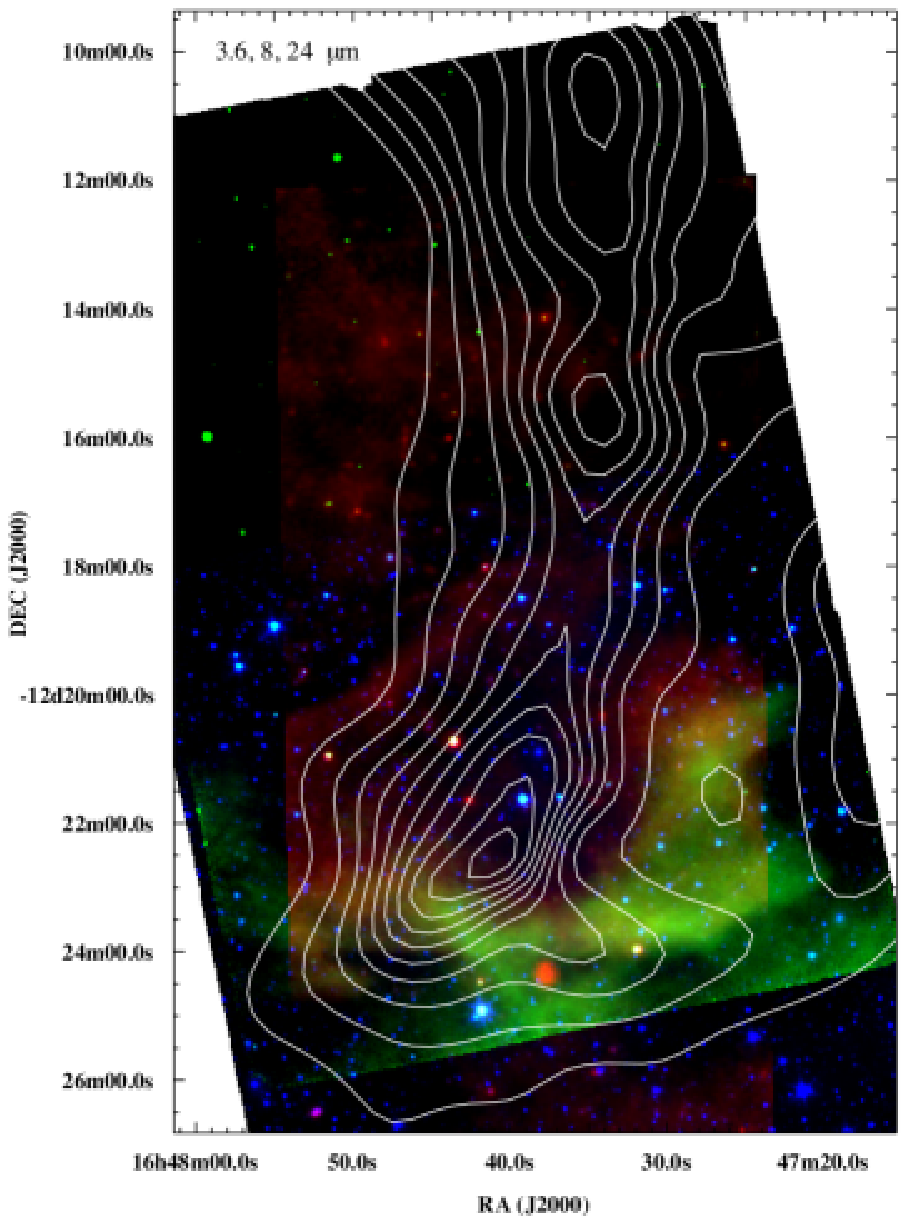}

\caption{\label{fig:rgb-l204c-2} Three-color image of L204C-2. The contours are
from the extinction map (\S\,\ref{sec:core-ext}). The $24$, $8$, and $3.6\:\mu$m
channels are used for the red, green, and blue emission, respectively.}

\end{figure}

\begin{figure}
\plotone{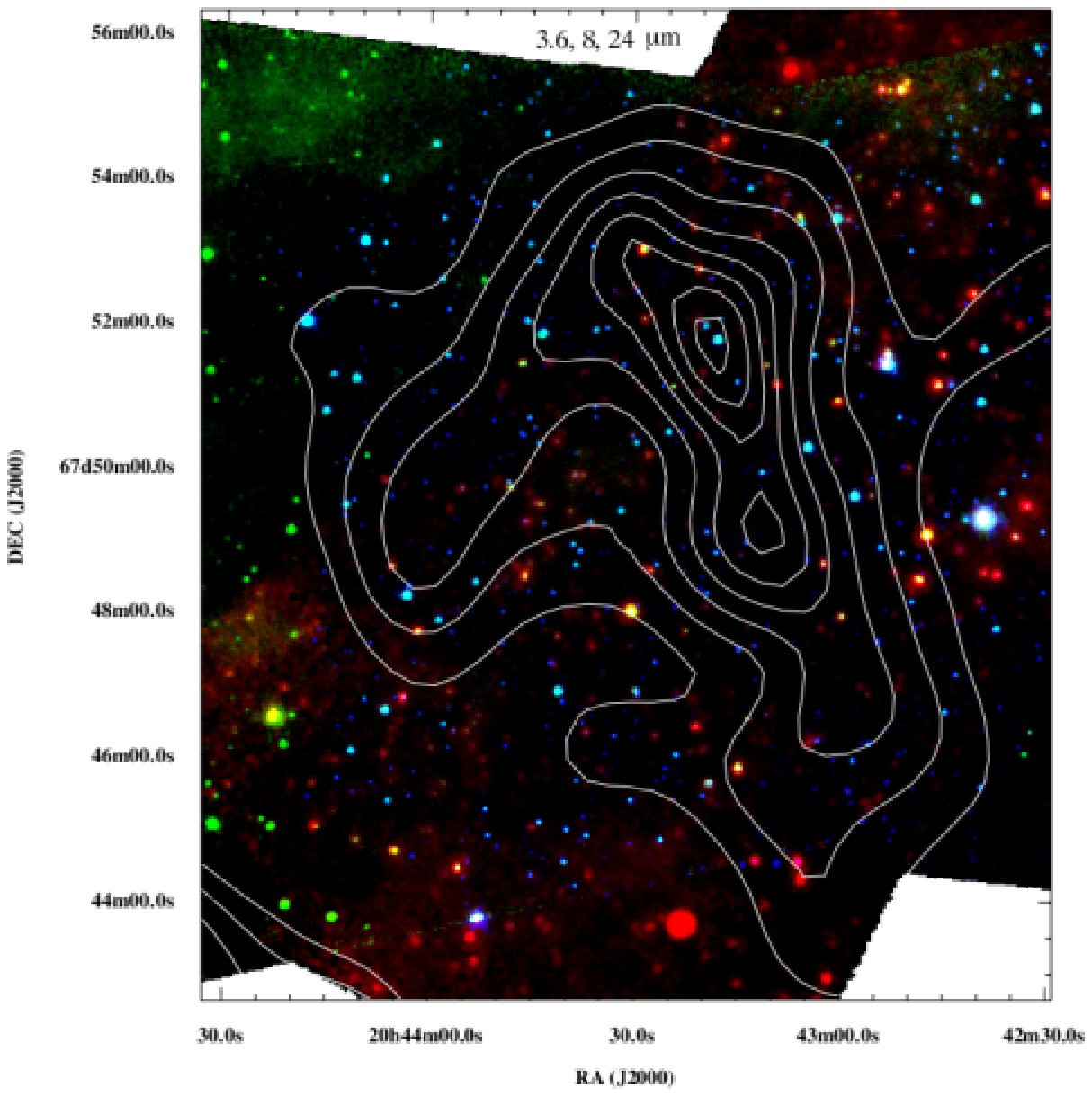}

\caption{\label{fig:rgb-l1155c-2} Three-color image of L1155C-2. The contours are
from the extinction map (\S\,\ref{sec:core-ext}). The $24$, $8$, and $3.6\:\mu$m
channels are used for the red, green, and blue emission, respectively.}

\end{figure}

To complement our \spitzer{} data, we obtained deep \jhks{} data using the
FLoridA Multi-object Imaging Near-ir Grism Observational Spectrometer
(FLAMINGOS) \citep{elston98} instrument installed on the 4-meter telescope at
Kitt Peak. These \jhks{} observations spanned four epochs from October 2003 to
June 2006. The FLAMINGOS field-of-view is $\sim11\arcmin \times 11\arcmin$. Each
field consisted of multiple dithers with small offsets around a central
position.  The total integration time per pixel was 12 minutes ($J$),  6 minutes
($H$), and 3 minutes ($K_s$).  For two fields each in L1152 and L1155C-2
(centered on the core) we took even deeper observations  with total integration
times of $\sim 1$ hour ($J$), $\sim 1$ hour ($H$) and $\sim45$ minutes ($K_s$).
Except for a small corner in L204C-2, the \jhks{} data completely cover the
areas mapped with \spitzer{}. In Figures
\ref{fig:rgb-l204c-2}-\ref{fig:rgb-l1228} we show three-color images of each
core and also selected subregions.

\subsection{ \label{sec:coredatareduce} Data Reduction }

We reduced our \jhks{} data using python and PyRAF. Each FLAMINGOS field
consisted of multiple dithers with small offsets around a central position. To
create our science images, we first subtracted a median filtered dark image from
each dither. Second, the dithers were flat-fielded by dividing out a dome
flat. Next, we subtracted off the sky. This is critical for IR observations
where the sky is typically much brighter than the astronomical objects of
interest. We used a two-pass sky subtraction with a running median: stars
were identified from the first pass, then masked out in the second
pass to produce an improved sky subtraction.  Finally, we averaged together the
individual dithers for each field.

We used the 2MASS catalog to correct the coordinates in each field and also to
calibrate the photometry. The stars in each field were found using the IRAF task
\texttt{daofind} and we used \texttt{daophot} to compute their point spread
function (PSF) photometry. We visually inspected the results to remove false
sources identified by \texttt{daofind} and also to add sources missed by the
source extractor. Finally, we bandmerged the individual $J$, $H$, and $K_s$
detections into a single \jhks{} catalog for each core.

Our \spitzer{} data were processed through the standard c2d pipeline.  For a
full description of this pipeline, please refer to the c2d data delivery
documentation available on the Spitzer Science Center's (SSC)
website\footnote{\url{http://ssc.spitzer.caltech.edu/legacy/c2dhistory.html}}.
We will summarize the pipeline here.  We started with the Basic Calibrated
Datasets (BCDs) processed by the SSC using their S13 pipeline.  We then
corrected several image artifacts caused by bright sources and cosmic rays and
also some instrumental effects.  Lastly, we applied location-dependent
photometric corrections that account for variations in the detector response
across the array.

After improving the BCDs we combined them into mosaics for each core using the
SSC's software MOPEX \citep{makovoz05}. We then extracted sources using a
modified version of DoPHOT \citep{schechter93}. DoPHOT fits PSFs and as sources
are found they are subtracted from the images. Next, we bandmerged the
individual detections from the $3.6$ to $24\:\mu$m wavebands and attempted to
determine upper limits for non-detections. Since we did not obtain any short
integration ``high dynamic range'' (HDR) observations, we used the c2d
observations that \emph{do} contain HDR frames to correct for saturated fluxes.
We compared our fluxes with the c2d fluxes in each IRAC band and empirically
determined a flux limit at which our values systematically diverged from c2d's.
We then substituted the c2d fluxes for any sources with a c2d flux brighter than
this limit. These limits were: 35, 35, 300, and 150 mJy for the 3.6, 4.5, 5.8,
and $8.0\:\mu$m bands, respectively.

Finally, we combined our $3.6$ to $24\:\mu$m and deep \jhks{} catalogs to create
bandmerged catalogs from $1.25$ to $24\:\mu$m for each of the four cores.

\subsection{ \label{sec:reliability} Data Quality}

The mean difference in position between sources in our \jhks{} catalogs and in
the 2MASS catalogs is $0\farcs15$, and 95\% of our sources have a difference in
position of $\leq 0\farcs6$ (around 2 pixels on the FLAMINGOS CCD). We also
computed the flux difference between our Kitt Peak observations and the 2MASS
catalog. The resultant distributions are gaussian with mean  $\lesssim 0.01$ mJy
and $\sigma = 4-5$\% in all three of the \jhks{} bands.

Even though our \spitzer{} data are not part of the c2d program, we processed
them through the c2d pipeline so we expect the errors in the \spitzer{}
wavebands will be similar to c2d's. The c2d delivery documentation discusses
three sources of uncertainty: statistical, systematic, and absolute. The first
two of these are incorporated into the photometric uncertainties listed in the
c2d catalog. These errors are derived from the repeatability of flux
measurements using the c2d pipeline. The c2d documentation lists a $4.6\%$
systematic error for IRAC and $9.2\%$ for MIPS. Combining the systematic and
statistical errors, our final photometric uncertainties are approximately $5\%$
in IRAC and $10\%$ in MIPS. The absolute uncertainties in the flux calibration
are 1.5\% and 4\%, respectively, for the IRAC and MIPS1 ($24\:\mu$m) bands. We
obtained these values from the Infrared Array Camera (IRAC) Data Handbook,
Version 3.0 and the Multiband Imaging Photometer for Spitzer (MIPS) Data
Handbook, version 3.3.0. The absolute uncertainties are added in quadrature with
the photometric errors during source classification, but are not used elsewhere
since they are smaller than the photometric errors.

Table \ref{tab:limits} lists our $10\sigma$ and $5 \sigma$ limits for each
wavelength. Our $10\sigma$ \jhks{} limits are $\sim3.5$ magnitudes deeper than
the 2MASS limits of 15.8, 15.1, and 14.3 magnitudes, respectively. In the dense
core regions of L1152 and L1155C-2  we were able to take even deeper \jhks{}
observations. These limits are listed in parentheses in the table. As for the
\spitzer{} wavebands, our $10\sigma$ limits are 1.5-2 magnitudes fainter in the
IRAC bands and 2.5 magnitudes fainter in the MIPS1 band compared to c2d.

\subsection{\label{sec:class} Extinction and Star Classification}

We used a modified version of the c2d data pipeline to classify sources based on
their spectral energy distribution (SED) from $1.25$ to $24\:\mu$m. The standard
c2d pipeline uses all available wavelengths to compute the extinction towards
each star:

\begin{equation}
\label{eq:fit}
\log(F_{obs}(\lambda)/F_{model}(\lambda)) =
\log(k)-0.4\times C_{ext}(\lambda) \times A_V
\end{equation}

\noindent where $F_{model}(\lambda)$ is the stellar photosphere model, $k$ is
the scaling factor of the model for a particular star, and
\mbox{$C_{ext}(\lambda) \equiv A_{\lambda}/A_V$} is the ratio of the extinction
at wavelength $\lambda$ to visual extinction from the dust extinction law. $k$
and $A_V$ can be derived from the linear \chisq{} fit of this equation by
adopting stellar photosphere and dust extinction models. The stellar models used
are based on the Kurucz-Lejeune models and come from the SSC's online
``Star-Pet'' tool\footnote{\url{http://ssc.spitzer.caltech.edu/tools/starpet}}.
For the 2MASS bands, the observed $J-H$ and $H-K$ colors of stars
\citep{koornneef83} were translated to fluxes relative to $K$ band and the
difference between the $K$ and $K_s$ bands was ignored.

Sources fit by Equation \ref{eq:fit} with a goodness-of-fit $q \ge 0.1$ were
classified as reddened stars, while those that did not fit were compared with
other templates to classify them.  The goodness-of-fit, $q$, is the incomplete
gamma function and gives the probability that the statistical \chisq{}
distribution would exceed  our observed \chisq{} (computed from Eq.\
\ref{eq:fit}) by chance.  Thus, a very small $q$ means it is unlikely that the
differences between our data and a  given $F_{model}$ are due to chance, i.e.\
the model is unlikely to be an accurate fit to the data \citep{press92}.  Our
minimum $q$ value of 0.1 is the same one used by the c2d team for identifying
stars.

We slightly modified this procedure for this paper.  First, the extinction is
computed for each source using the \jhks{} bands and the NICER technique
\citep{lombardi01}.  The NICER technique relies on the assumption of intrinsic
values for the $J-H$ and $H-K_s$ colors of stars.  For this paper we adopted the
values $J-H^{intrinsic} = 0.50\pm0.12$ and ${H-K_s}^{intrinsic} = 0.18\pm0.04$.
These values were derived from the average stellar model computed in Appendix
\ref{sec:avgmodel}.  Then, using Equation \ref{eq:fit}, we identified the stars
with the extinction held fixed.  We also computed extinctions in \ak{} rather than
the more traditional $A_V$ so we can directly compare our results with those of
other authors.

We made this change because one of our goals in this paper is to compute the
extinction law at the IRAC and MIPS wavelengths. The method we use for deriving
the mid-infrared extinction law (\alak) is dependent on \ak{} so we cannot use
these wavelengths when computing \ak. However, we still need to choose an
appropriate extinction law. This will be used to compute \ak{} from the \jhks{}
bands and also for the source classification. We considered two dust models, the
\citet{weingartner01} $R_V = 3.1$ and $R_V = 5.5$ models. The $R_V = 3.1$ model
(hereafter WD3.1) is designed to reproduce the extinction law of the diffuse
interstellar medium; the $R_V = 5.5$ model (hereafter WD5.5) fits the observed
law of denser regions. We found in \citet{chapman09} (hereafter \paperone) that
some heavily extincted stars may only fit the WD5.5 dust model rather than the
more traditional WD3.1 extinction law. In lower extinction regions both models
tend to fit equally well. Therefore, for source classification we have chosen to
use the WD5.5 dust model.

\subsection{ \label{sec:core-highreliability} High Reliability Star Catalogs }

The c2d data pipeline classifies sources according to their SED and infrared
colors. Among other classifications, sources are identified as stars, `Galaxy
candidates' (Galc), or `Young Stellar Object candidates' (YSOc). With YSOc,
Galc, and other infrared excess sources we cannot separate out the expected
flux of the SED from changes in those fluxes due to variations in the extinction
law. Therefore, we want to concentrate on those sources classified as stars for
studying the variations in the extinction law.

To create our high-reliability star catalogs we first selected all sources
classified as stars, excluding those classified as stars when one waveband was
dropped because we did not want a prominent absorption or emission line to bias
our results. Furthermore, because accurate extinctions are essential to our
analysis, we required sources to have detections $\ge 7\sigma$ in each of the
$JHK_s$ bands. Lastly, we removed suspected faint background galaxies from our
star catalogs using the criteria listed in \S\,\ref{sec:cores-faintgalaxy}.

The number of stars in our high-reliability catalogs is: 3,244 (L204C-2), 4,038
(L1152), 3,926 (L1155C-2), and 1,850 (L1228).  We then removed a few sources, as
described below, to obtain our final catalogs.  The percentage of sources
removed is $< 1\%$ for L204C-2, L1152, and L1155C-2, and 2.2\% in L1228.

\subsection{\label{sec:cores-faintgalaxy} Misidentified Background Galaxies}

\begin{figure}

\plotone{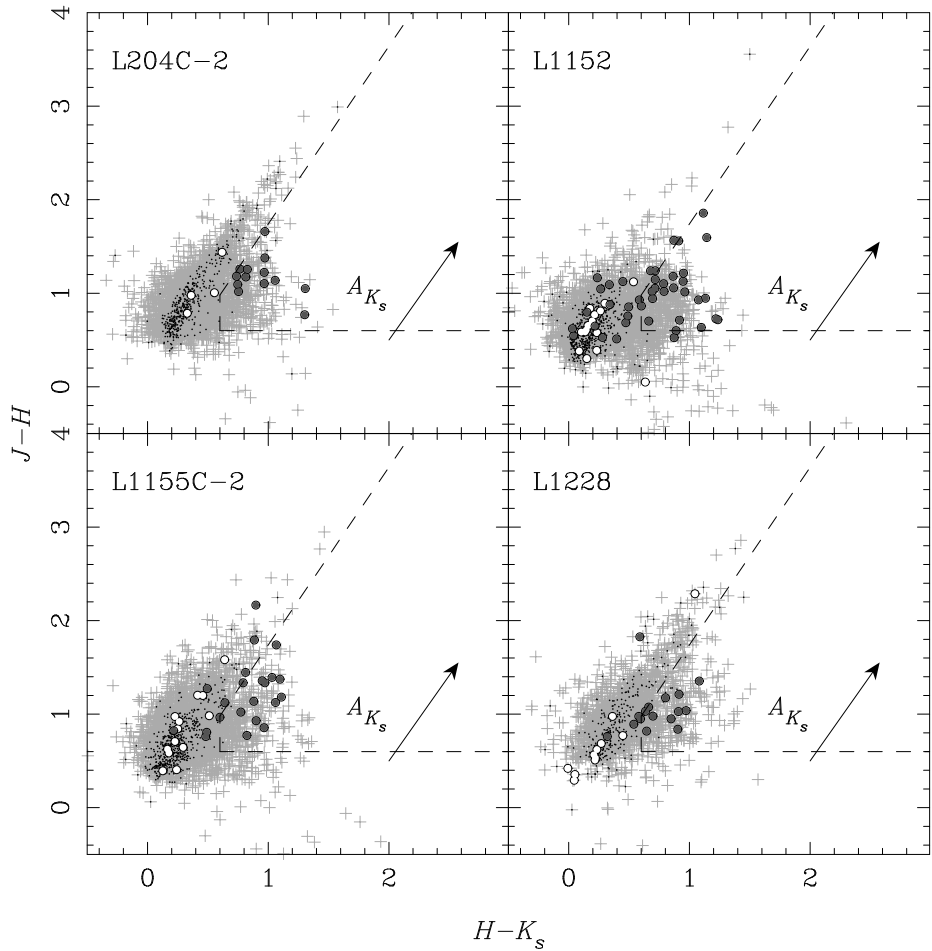}

\caption{\label{fig:ahak-withgal} $J-H$ vs.\ $H-K_s$ color-color diagram of the
high-reliability stars for each core.  The gray crosses are all the stars, while
the black  points are those stars brighter than 15th magnitude at $K_{s}$.
Overlaid on the stars are white and dark gray circles, which are the `known'
stars and background galaxies, respectively. These sources were selected from
Figure \ref{fig:corek24}. The reddening vector for our extinction law, WD5.5, is
also shown.  See \S\,\ref{sec:cores-faintgalaxy} for a description of the dashed
lines and further details.}

\end{figure}

Although we carefully selected only the most reliable stars to construct our
catalogs, it appears that some faint background galaxies are misidentified as
stars.  This can be readily seen in Figure \ref{fig:ahak-withgal}, where we plot
$J-H$ versus $H-K_s$ for the stars in our cores.  The stars brighter than 15th
magnitudes at $K_s$ (black points) follow the reddening vector, while many of
the fainter sources (gray crosses) do not.  The `known' stars and background
galaxies are shown as white and dark gray circles, respectively. Our method for
creating these two populations of sources is described below.  Most of the
`known' background galaxies are grouped to the right of the bright sources, in
the same area as those gray crosses which do not follow the reddening vector.
This suggests that these gray crosses are actually background galaxies.
\citet{foster08}, using near-infrared data from Perseus, also concluded that
sources in this approximate region of a $J-H$ versus $H-K_s$ color-color diagram
are likely to be background galaxies.  Background galaxies are not the only
explanation for these sources.  It is possible that a few of them may be YSOs.
In \S\,\ref{sec:constraints} we will discuss the prospects for faint YSOs in
the cores.

To create samples of `known' stars and background galaxies, we could not start
from the high-reliability catalogs since those explicitly exclude non-star
objects.  Instead, we began with the full catalogs and imposed the same
$7\sigma$ cutoff in the \jhks{} bands as we did for the high-reliability star
catalogs.  Second, we selected only those sources with $24\:\mu$m detections
$\ge3\sigma$.  The resultant sources are plotted in Figure \ref{fig:corek24}. We
used different symbols depending on how sources were classified in
\S\,\ref{sec:class}: stars are shown as circles, Galc sources are squares,
YSOc's are shown as triangles, and plus signs for everything else. Furthermore,
the shaded contours are the c2d processed SWIRE data of region ELAIS N1
\citep{surace04}. Since the ELAIS N1 region is near the north Galactic pole, it
should contain nothing but stars and background galaxies making it useful for
comparison. From this figure, we selected `known' populations of stars and
background galaxies. The stars are those sources with $K_s - [24] \leq 1$ and
the background galaxies have $K_s \geq 15$ and $K_s - [24] \geq 4$.

\begin{figure}

\plotone{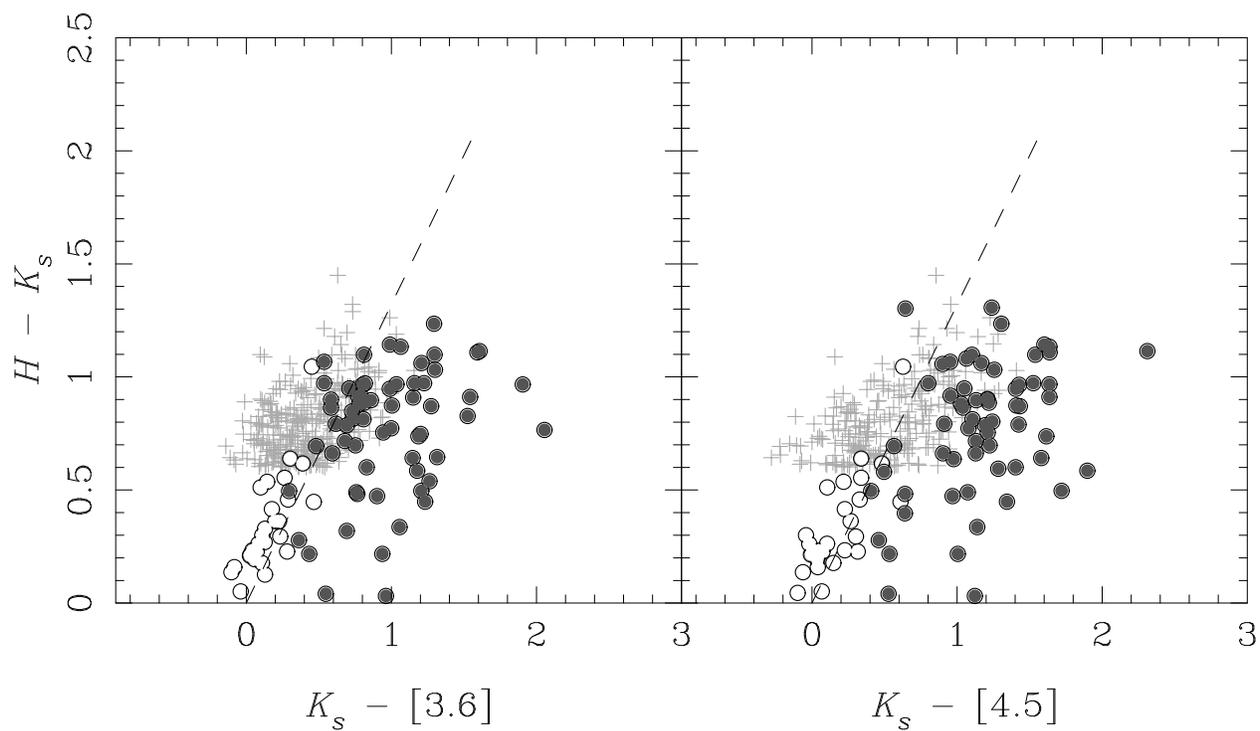}

\caption{\label{fig:cc-bad-cores} $H-K_s$ vs.\ $K_s - [3.6]$ (left) and vs.\
$K_s - [4.5]$ (right). The gray crosses are the sources selected by the dashed
lines in Figure \ref{fig:ahak-withgal} while the white and dark gray
circles are the `known' stars and background galaxies, respectively. The dashed
line in each panel has the equation $H-K_s = 1.32x$ where $x$ is either $K_s -
[3.6]$ or $K_s - [4.5]$.}

\end{figure}

We will follow a procedure similar to that in \paperone{} to eliminate the
misidentified background galaxies from our high-reliability star catalogs.
First, we select suspected background galaxies with the following colors: $J-H
\ge 0.6$, $H-K_s \ge 0.6$, and $J-H \le 1.9\times (H-K_s) - 0.16$. These
selections are shown as dashed lines in Figure \ref{fig:ahak-withgal}. Then,
just using the sources selected by these colors, we plotted them on two
color-color diagrams in Figure \ref{fig:cc-bad-cores} along with our `known'
star and background galaxy samples.  There is a clear separation between the
`known' stars and background galaxies. The dashed line in the figure is: $H-K_s
= 1.32x$, where x is either $K_s - [3.6]$ or $K_s - [4.5]$. We combine these two
selections to remove our misidentified background galaxies. These selection
criteria are the same empirically derived ones used in \paperone{}, but with one
important difference: because there is such a large degree of non-overlap
spatially between the four IRAC bands, we relaxed the second criterion from
\paperone{} so that sources only need to satisfy either the first half or the
second half of it. Our criteria are:

\begin{enumerate}

\item $J-H \ge 0.6$, $H-K_s \ge 0.6$, and $J-H \le 1.9 \times (H-K_s) - 0.16$

\item $H-K_s \le 1.32 \times (K_s - [3.6])$ or $H-K_s \le 1.32 \times (K_s - [4.5])$

\end{enumerate}

Sources satisfying both criteria were removed from our catalogs.  After removal
of these sources, The number of stars in our final high-reliability catalogs is:
3,219 (L204C-2), 4,010 (L1152), 3,891 (L1155C-2), and 1,810 (L1228).

\section{\label{sec:core-ext} Extinction Maps}

Our star catalogs contain a line-of-sight extinction measurement to each star
calculated using the NICER technique \citep{lombardi01} and the WD5.5 dust
model. We converted these randomly distributed samplings of the true extinction
within each core into uniformly sampled maps by overlaying a grid on each core
with an $18\arcsec$ spacing between grid elements. At each grid position, an
extinction value was computed as the average of the individual measurements
within a $90\arcsec$ radius. This average is weighted both by the uncertainty of
each line-of-sight extinction value and also by a gaussian function with full
width half maximum (FWHM) equal to $90\arcsec$. Thus, our final maps have
$90\arcsec$ resolution with 5 pixels across the FWHM.  By setting our
integration radius to equal the FWHM ($2.3548\sigma$), we capture $98.14\%$ of
the total area of the gaussian weighting function, and avoid the unrealistic
scenario of utilizing extinction values from arbitrarily large distances away
from each grid position.

We subtracted foreground stars in a simple way: For each cell with more than 2
stars, we computed the mean and median \ak{} values. If these two statistics
differed by $>25\%$, then we dropped the source with the lowest \ak{} and
recomputed the mean. If the difference between the old and new mean is less than
25\%, we re-added the dropped source to our catalog. After identifying all
sources to drop, we recomputed the extinctions in every cell.  Despite its
simpleness, we found this procedure to be very effective in removing foreground
stars.  The $25\%$ cutoff was empirically chosen after we ran some tests 
with different cutoff parameters.

\begin{figure}
\epsscale{0.87}
\plotone{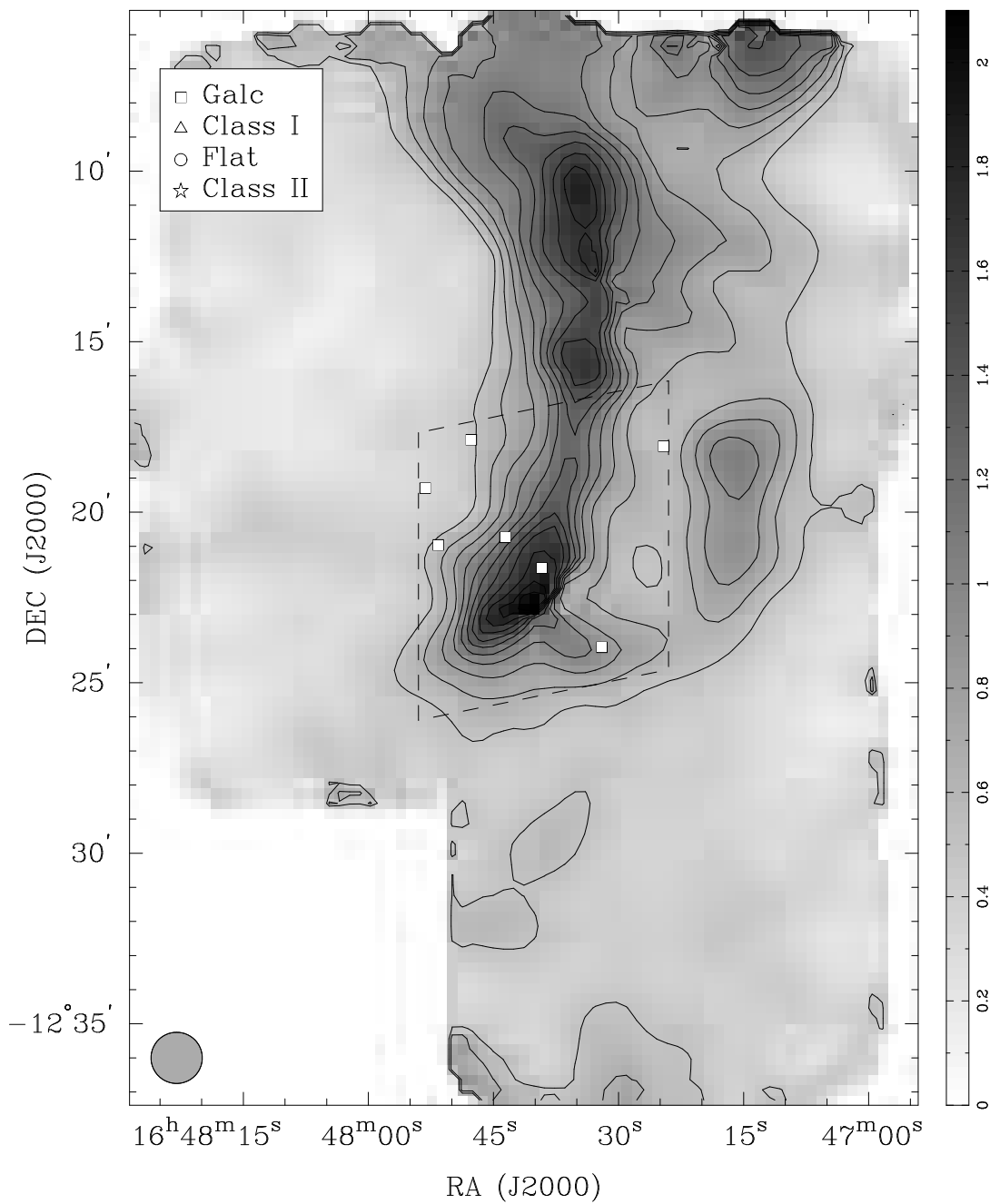}

\caption{\label{fig:l204c-2ak} Extinction map of L204C-2. The map has a
resolution of $90\arcsec$ with a maximum \ak{} value  of 2.1 magnitudes.  The
contours start at $A_{K_s} = 0.5$ mag.\ in steps of 0.15 ($3\sigma$).  Sources
classified as Galc are shown as squares, while YSOs are shown with one of three
symbols, depending on their evolutionary classification \citep{greene94}.  There
are no YSOs plotted in this figure.  The dashed region denotes the approximate
area covered with both IRAC and MIPS.  The gray circle denotes the resolution
of the map.}

\end{figure}

\begin{figure}
\epsscale{1}
\plotone{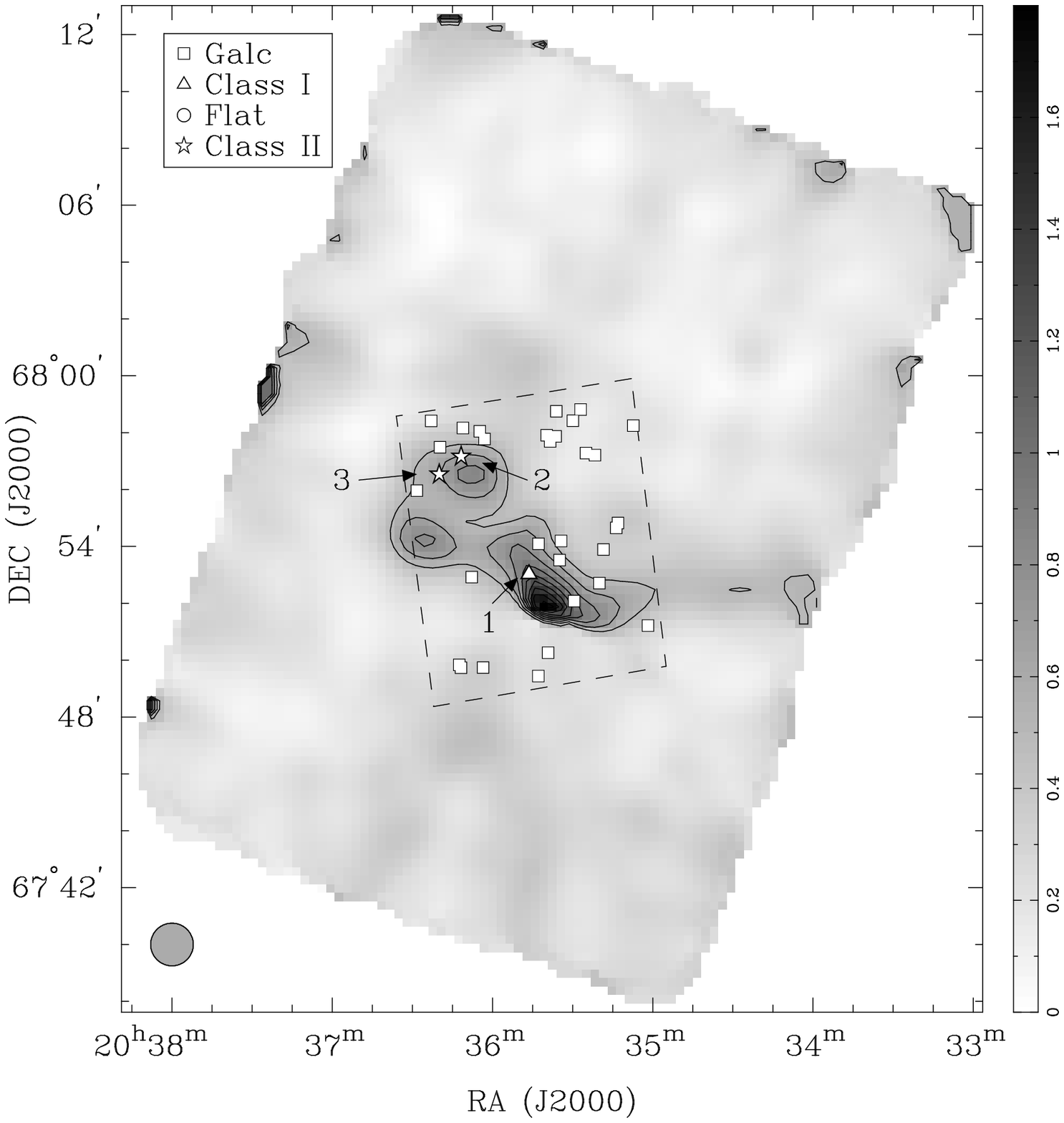}

\caption{\label{fig:l1152ak} Extinction map of L1152.  The  maximum \ak{} value
in the map is 1.8 magnitudes.  See Figure \ref{fig:l204c-2ak} for a  description
of the contours, symbols, and dashed lines.  The numbered sources are the YSOs
discussed in \S\,\ref{sec:coreysos}. }
\end{figure}

\begin{figure}

\plotone{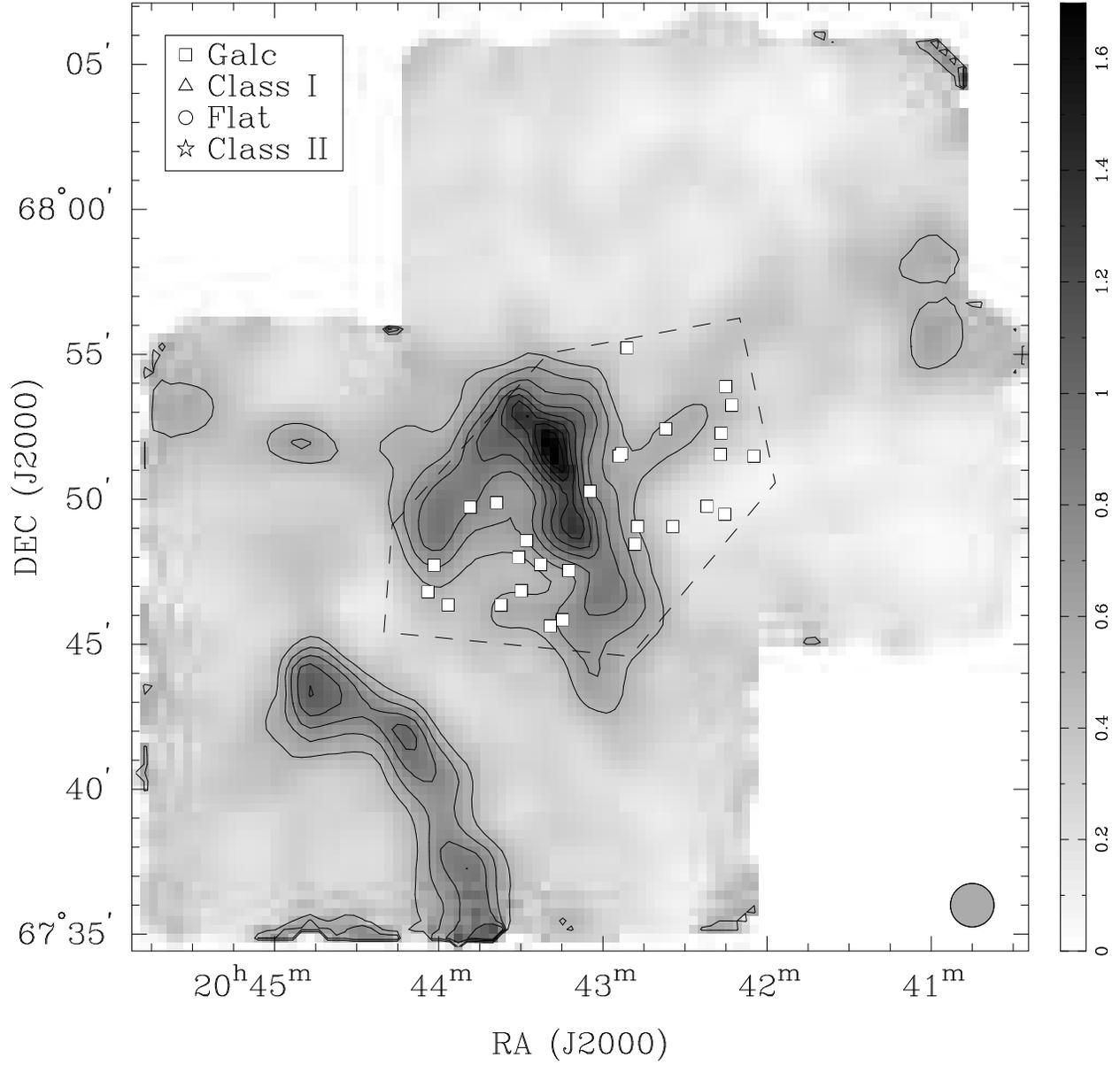}

\caption{\label{fig:l1155c-2ak} Extinction map of L1155C-2.  The maximum \ak{}
value in the map is 1.7 magnitudes.  See Figure \ref{fig:l204c-2ak} for a
description of the contours, symbols, and dashed lines.  There are no YSOs
plotted in this figure.}

\end{figure}

\begin{figure}

\plotone{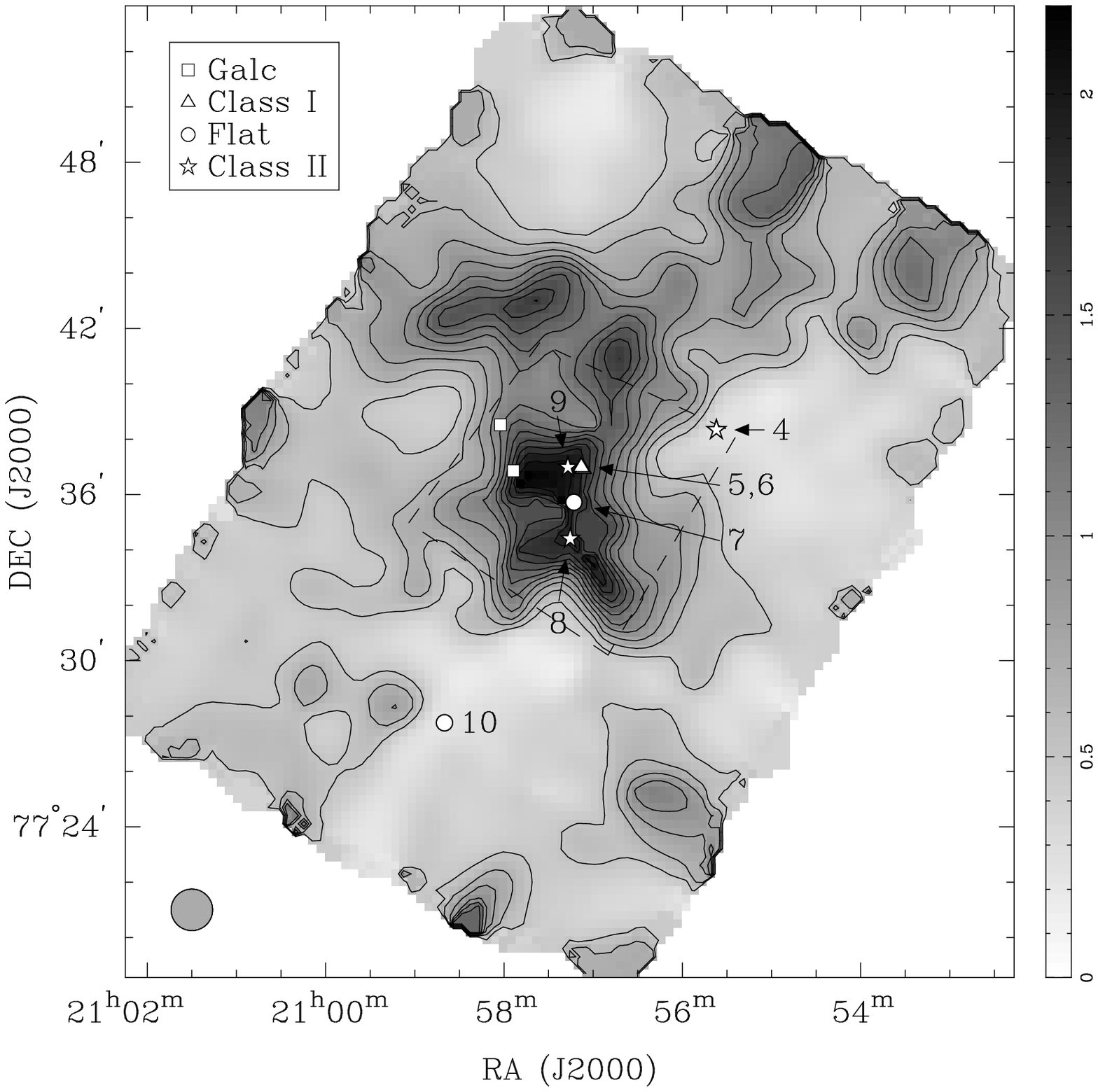}

\caption{\label{fig:l1228ak} Extinction map of L1228. The  maximum \ak{} value
in the map is 2.2 magnitudes.  See Figure \ref{fig:l204c-2ak} for a  description
of the contours, symbols, and dashed lines.  The numbered sources are the YSOs
discussed in \S\,\ref{sec:coreysos}.}

\end{figure}

We chose the $90\arcsec$ resolution because this gave us a reasonable number of
stars per cell for good statistics. The median number of stars per cell is
$\sim35-40$ for L204C-2, L1152, and L1155C-2, and 18 in L1228. The percentage of
cells with $\leq 5$ stars is $6$\% in L1152 and L1155C-2, $11$\% in L204C-2, and
$14$\% in L1228. Figures \ref{fig:l204c-2ak} - \ref{fig:l1228ak} are extinction
maps of our four cores. The contours in each map start at $A_{K_s} = 0.5$ mag in
steps of 0.15 ($3\sigma$). We also plotted the YSOc and Galc sources from our
catalogs on the extinction maps and numbered the YSOs from
\S\,\ref{sec:coreysos}. Lastly, because YSOc and Galc sources can be classified
only in regions with IRAC1 through IRAC4 and MIPS1, we outlined the area in each
core where all these bands overlap.

The core of L204C-2 has a peak extinction of $A_{K_s} = 2.1$. Extending
northward from the central core is a dust filament with two other lower density
extinction peaks. West of the core is a low density clump with a peak extinction
of $A_{K_s} = 0.8$. We detect no YSOs in L204C-2. One Galc object is located
near the center of the core, however, as we will discuss in
\S\,\ref{sec:coreysos} this source has an SED consistent with a heavily
extincted background star.

L1152 is a small core with peak $A_{K_s} = 1.8$. To the east are two additional
separate extinction peaks. Both of these have a lower column density than the
main core. A bright YSOc is embedded in the primary core, slightly northeast of
the extinction peak. One lobe of an outflow can be seen in $J$ through IRAC4
extending southwest from this YSO. The other two YSOs are northeast of the core,
and slightly offset from the northernmost extinction peak.

L1155C-2 is an extended core with two extinction peaks, the northernmost has a
higher peak extinction of 1.7 magnitudes. A dust filament that is part of L1155E
appears in the southern third of the map. Like L204C-2, this core also appears
starless.

The central core of L1228 contains two extinction peaks and appears pinched in
the middle. The peak extinction in this core is 2.2 magnitudes. This pinching
may simply be an artifact caused by the lower median number of stars per cell
compared with the other cores or it may indicate clearing by the outflow since
the pinch direction lines up with the east-west outflow seen in Figure
\ref{fig:rgb-l1228}. Several other extinction peaks are nearby, mostly north of
the core. L1228 has previously been classified as a starred core, and several
YSOs are identified in the core; most have not been previously identified.
Despite having similar Galactic coordinates as L1152 and L1155C-2, only two Galc
objects appear in this core. This is because L1152 and L1155C-2 have deeper
\jhks{} data than L1228, therefore many more Galc objects fall into our
high-reliability catalogs for those cores than in L1228.

\section{ \label{sec:coreysos} Young Stellar Objects }

\begin{figure}
\epsscale{1}
\plotone{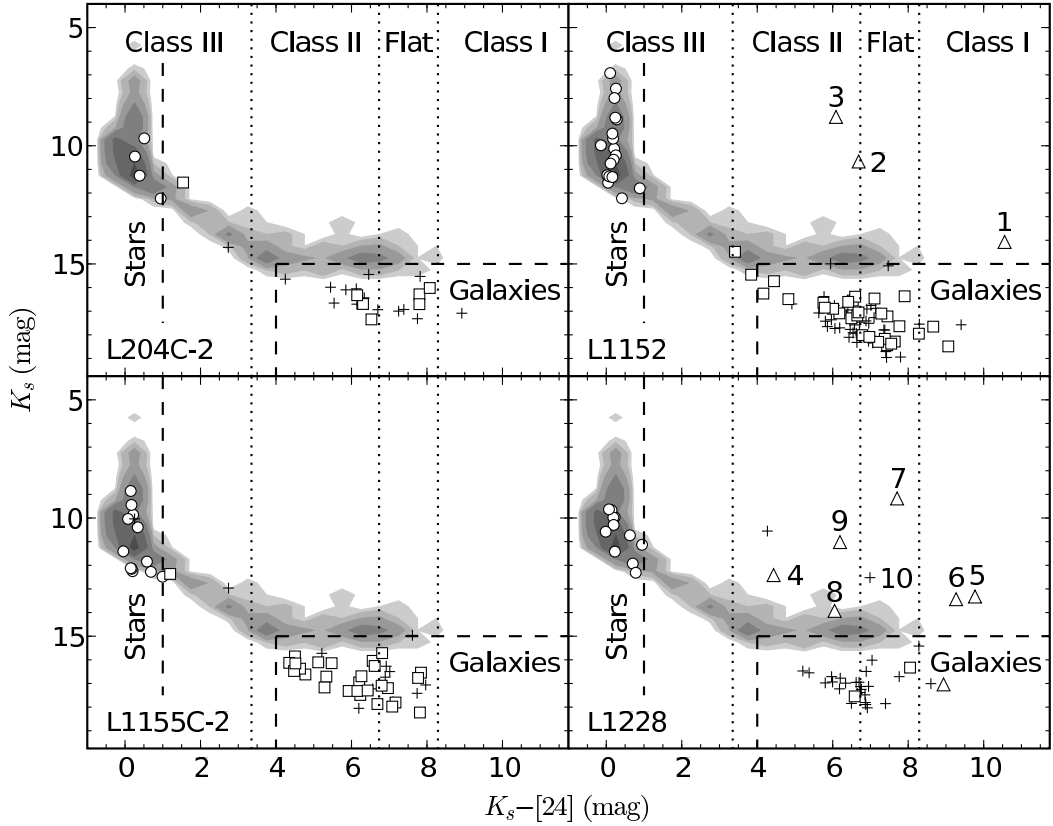}

\caption{\label{fig:corek24} $K_s$ vs. $K_s - [24]$ plot of all sources with
\jhks{} detections $\ge 7\sigma$ and $24\:\mu$m $\ge 3\sigma$. We used four
different symbols to correspond to different source classifications: stars are
circles, Galc sources are squares, YSOc objects are triangles, and plus signs
denote all other classifications.  The shaded contours are from the c2d
processed catalog of part of ELAIS N1 \citep{surace04}.  The dashed lines denote
the cutoffs we used for selecting `known' stars and background galaxies, while
the dotted lines are the cutoffs for different YSO evolutionary classes as
defined by \citet{greene94}.  The YSOs discussed in this paper are  numbered for
easy reference.}

\end{figure}

Ten sources are classified as YSOc in our cores. However, a source can be
classified as a YSOc only if it is detected in all \spitzer{} wavebands from
$3.6$ to $24\:\mu$m. This is because various empirically derived \spitzer{}
colors and magnitudes are used in classifying YSOc sources \citep{harvey07}.
Therefore, to our initial list of 10 sources we added six potential YSOs that
were not classified as YSOc. First, we included sources from Figure
\ref{fig:corek24} with $K_s - [24] \ge 3$ and $K_s \le 14.5$. Most of the
sources in this region are classified as YSOs already. This selection added two
sources in L1228, neither of which is classified as YSOc because they are
outside of the area observed in IRAC2 and IRAC4 ($4.5$ and $8\:\mu$m). It is
also possible that faint YSOs may be misidentified as Galc. Therefore, we also
selected four Galc objects from Figures \ref{fig:l204c-2ak}-\ref{fig:l1228ak}
that were located near the dense cores; one source each in L204C-2 and L1152 and
two in L1228. This brought our total number of possible YSOs to 16.

After we visually inspected all 16 sources, we excluded the Galc objects in
L1152 and L1228 since all were visually extended in mosaics, meaning they are
likely to be true background galaxies. We also excluded the one Galc object from
L204C-2 because this source has the SED of a heavily extincted star. In Figure
\ref{fig:corek24}, this source is the Galc at $K_s - [24] = 1.5$. Furthermore,
we excluded one of the YSOc objects in L1228 that is clearly a knot of emission
from the outflow and not a real source. Lastly, we also excluded one of the
non-YSOc sources in L1228 that only has an infrared excess at $24\:\mu$m. The
$24\:\mu$m flux for this source appears confused with two other nearby sources
detected at shorter wavelengths. We are left with 10 YSOs, three in L1152 and
seven in L1228. Note that now we are identifying these sources as Young Stellar
Objects and not Young Stellar Object candidates. In Table \ref{tab:coreysos} we
list the fluxes and positions of these 10 YSOs present in our cores. For each
YSO, we added the $70\:\mu$m flux from c2d, if detected. Neither L204C-2 nor
L1155C-2 appear to contain any YSOs. In \S\,\ref{sec:constraints} we will place
some upper limits on unidentified YSOs in our cores.

\begin{figure}
\epsscale{0.48}
\plotone{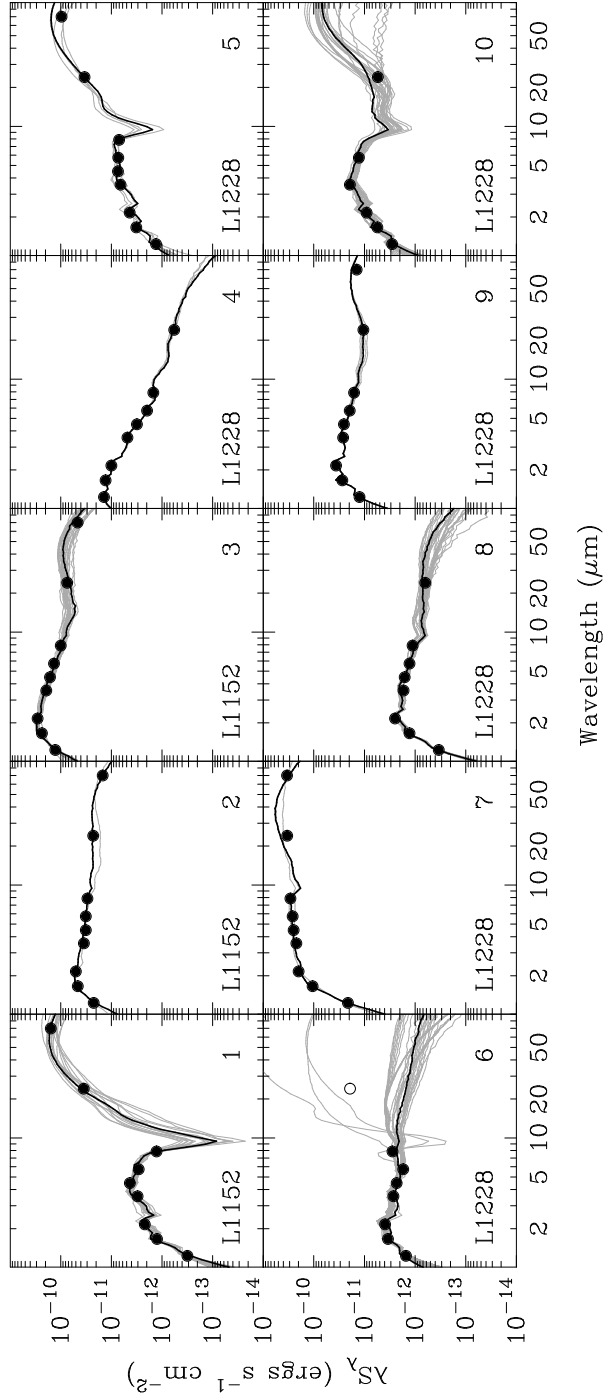}

\caption{\label{fig:yso-sed} The spectral energy distributions for our 10 YSOs.
The black curve is the best-fit YSO model from \citet{robitaille07} while the
gray curves show other model fits with $\chi^2 \leq 2\times \chi^2_{best}$
(\S\,\ref{sec:coreysos}). The errorbars are smaller than the datapoints. Each
YSO is numbered for reference. }

\end{figure}

We modeled the SED for our 10 YSOs using an online tool \citep{robitaille07}.
This tool fits an input SED to a pre-computed grid of YSO models.  In Figure
\ref{fig:yso-sed} we show all models with $\chi^2 \leq 2 \times \chi^2_{best}$,
where $\chi^2_{best}$ is the \chisq{} for the best-fitting model.  The best-fit
model for each YSO is shown in black with other models shown in gray. Each model
has numerous parameters, a few of which we have listed in Table
\ref{tab:robitaille}.  We show the range of values for envelope accretion rate,
disk mass, interstellar extinction ($A_V$), and bolometric luminosity plus the
average value for each quantity.  Note that the Robitaille models do include
stellar masses, however these are not constrained for embedded objects, so we
will not use them here (Robitaille, private communication).

We also classified our YSOs based on their value of $\alpha = d \log  \lambda
F_\lambda/d \log \lambda$, the best-fit slope of a straight line to the SED from
$K_s$ to $24\:\mu$m.  We then separated the YSOs into one of four classes using
the method of \citet{greene94}.  Following an evolutionary sequence from
youngest to oldest YSO, they are: Class I ($\alpha \ge 0.3$); Flat spectrum
($0.3 > \alpha \ge -0.3$); Class II ($-0.3 > \alpha \ge -1.6$); and Class III
($\alpha < -1.6$).  We have included sources with $\alpha = 0.3$ in Class I
since these were undefined by \citet{greene94}. Our YSOs are a mixture of Class
I, Flat spectrum, and Class II sources, but no Class III objects.  This suggests
that the YSO population in these cores is relatively young.

\subsection{L1152 YSOs}

\begin{figure}
\epsscale{1}
\plotone{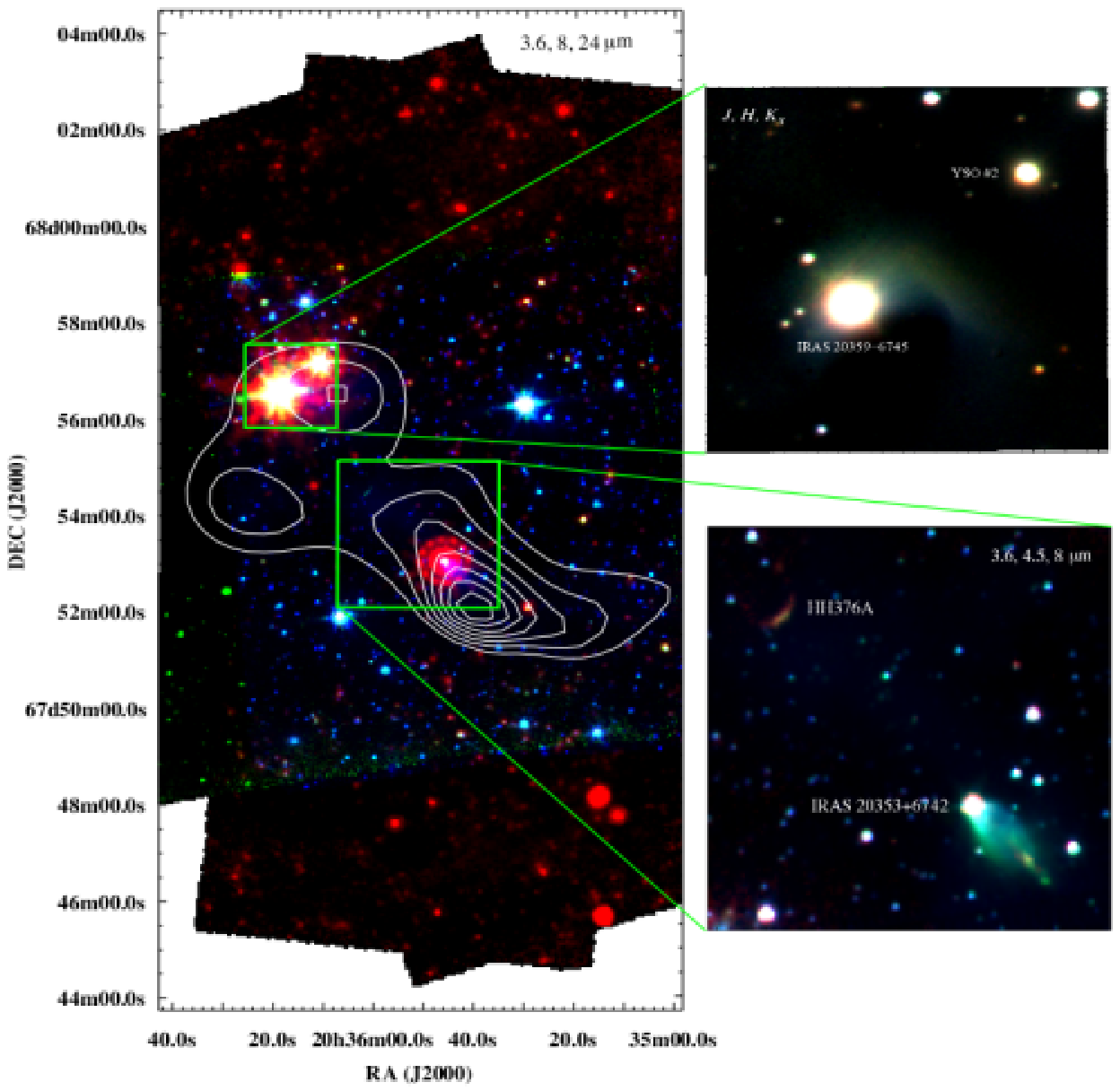}

\caption{\label{fig:rgb-l1152} Three-color image of L1152. The contours are from
the extinction map (\S\,\ref{sec:core-ext}). The wavelengths used for the  blue,
green, and red emission, respectively, are listed at the top of each
sub-figure.}

\end{figure}

Two of the three YSOs in L1152, IRAS 20353+6742 and IRAS 20359+6745, are
previously known and both have outflows associated with them. IRAS 20353+6742 is
the source embedded in the L1152 core and is source \#1 in Tables
\ref{tab:coreysos} and \ref{tab:robitaille}. Only one lobe of the outflow is
visible, but it can be seen from $J$ through IRAC4 (Figure \ref{fig:rgb-l1152}).
The second known source, IRAS 20359+6745, is source \#3 in Tables
\ref{tab:coreysos} and \ref{tab:robitaille}. This YSO has an outflow associated
with it that is visible in \jhks{} but not at longer wavelengths. The
Herbig-Haro object HH376A is located $2\farcm6$ to the southwest. The bowshock
shape of HH376A and its location along the apparent axis of the \jhks{} outflow
from source \#3 suggests it may be part of the same system. We show both of
these features in Figure \ref{fig:rgb-l1152}. The remaining YSO is approximately
$1\arcmin$ to the west of IRAS 20359+6745. We found no references to this source
in SIMBAD\footnote{http://simbad.u-strasbg.fr/simbad/}.

\subsection{L1228 YSOs}

\begin{figure}
\epsscale{0.74}
\plotone{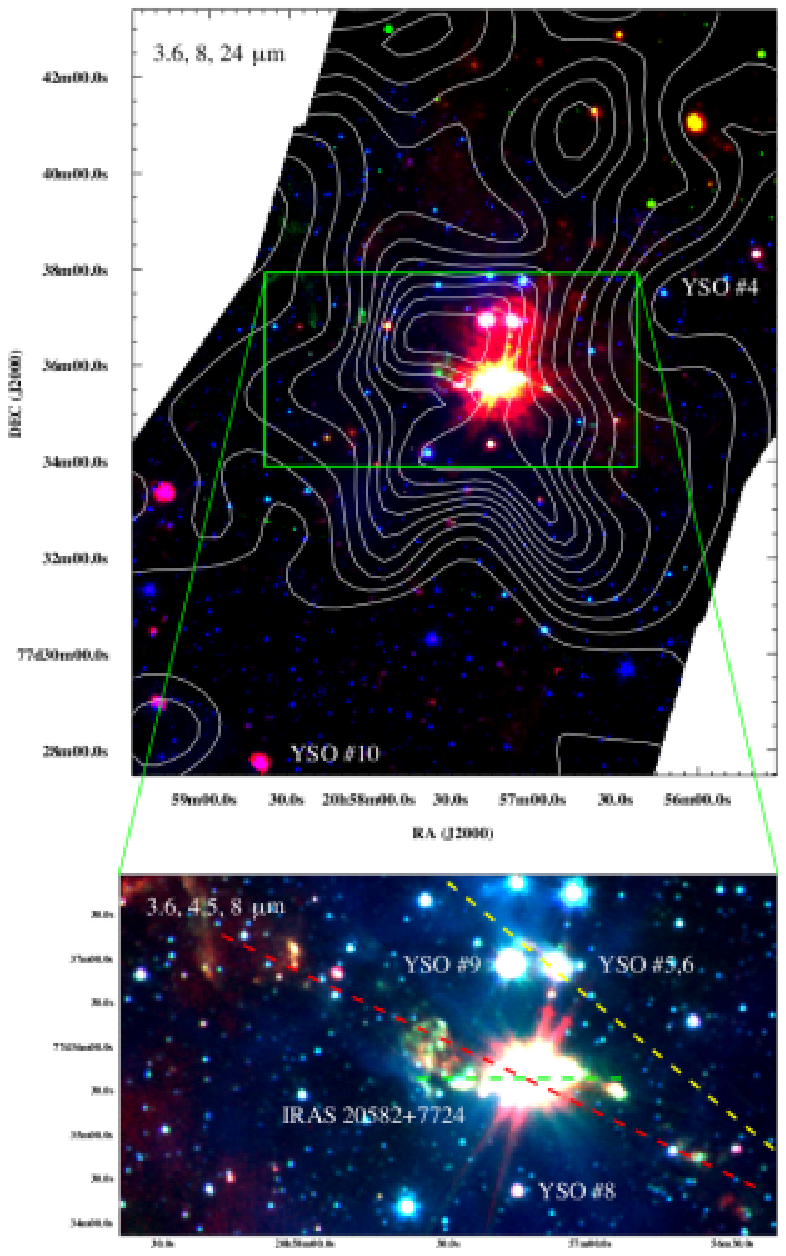}

\caption{\label{fig:rgb-l1228} Three-color image of L1228. The contours are from
the extinction map (\S\,\ref{sec:core-ext}). The wavelengths used for the blue,
green, and red emission, respectively, are listed at the top of each
sub-figure.}

\end{figure}

We identified seven YSOs in L1228. Six of these are classified as YSOc based on
their colors while the remaining one cannot be classified as such because it
appears outside of the area covered by IRAC2 and IRAC4. Five of the YSOc sources
appear to be spatially coincident with the core of L1228. The brightest source
is \#7 and is identified as IRAS 20582+7724. This is the only previously
known YSO in the core of L1228.  It appears to be the driving source
for a CO outflow, HH 199 \citep{haikala89,bally95}.  The axis of the CO outflow
is about $40^\circ$ different from the east-west axis formed by the infrared
knots.  \citet{bally95} conclude this is likely due to precession of the jet
that drives the outflow. In Figure \ref{fig:rgb-l1228} the green dashed line
shows the axis of the infrared knots while the red dashed line shows the axis of
the CO outflow.

Sources \#5 and \#6 are located about $1\farcm2$ northwest of IRAS 20582+7724
and separated from each other by $5\farcs3$.  Because of this separation, they
are not resolved at $70\:\mu$m ($FWHM = 17\arcsec$) and not fully resolved at
$24\:\mu$m ($FWHM = 5\farcs7$).  Even though we were able to extract a flux at
$24\:\mu$m for both sources, we suspect that the $24\:\mu$m flux for source \#6
is artificially high.  When fitting YSO models, we found that excluding this
datapoint decreased $\chi^2_{best}$ from 174 to 48, the latter is more in line
with the $\chi^2_{best}$ for the other sources.  Therefore, we excluded this
datapoint for the models listed in Table \ref{tab:robitaille} and shown in
Figure \ref{fig:yso-sed}.  It is also possible that the $70\:\mu$m flux for
source \#5 includes some contribution from source \#6 as well.  \citet{bally95}
detected a series of infrared emission knots that form the HH 200 outflow which
has its origin at the position of sources \#5 and \#6. It is unknown which of
these is the driving source of the outflow.  We show the approximate axis of
this outflow as a yellow dashed line in Figure \ref{fig:rgb-l1228}.  There is
some outflow emission immediately to the northeast of sources \#5 and \#6 which
lies along the outflow axis.

Lastly, the two YSOs not spatially coincident with the L1228 core are sources
\#4 and \#10.  Source \#10 does not meet the criteria for YSOc because it is
lacking IRAC2 and IRAC4 fluxes.

\subsection{\label{sec:constraints} Constraints On New Faint YSOs}

We did not detect any YSOc sources that were not already detected by c2d. The
converse is also true, in the c2d catalogs of all four cores, there is only one
YSOc object not in our catalogs. However, after inspection we determined this
object to be another emission knot from the east-west outflow. This lack of
additional YSOc sources is not too surprising since the YSOc selection criteria
were tuned to the sensitivity of the c2d-processed SWIRE data. Therefore, any
sources fainter than those limits are likely to be automatically classified as
Galc.  It was for this reason that we also considered some Galc sources when
compiling our YSO catalog. However, from Figures
\ref{fig:rgb-l204c-2}-\ref{fig:rgb-l1228} it is clear that there are many bright
$24\:\mu$m sources in our cores, far more than the number of YSOc and Galc
objects.

We would like to use these objects to place limits on possible faint YSOs in
these cores. If we select sources from the overlap region of all five bands with
$24\:\mu$m detections $\ge 3\sigma$, then exclude any star or YSOc sources and
also any Galc sources in Figure \ref{fig:corek24}, we are left with 634 sources
in all four cores. These 634 sources have a median $24\:\mu$m flux of 0.21 mJy
(11.3 magnitude). Just 47 of them are detected at $K_s$, but these have a median
flux of 0.04 mJy (18.1 magnitude). Our $5\sigma$ detection limit for $K_s$ is
18.4 magnitudes, or 19.3 magnitudes for the deep observations. If we assume that
the remaining sources must be fainter than these limits in order to be
undetected at $K_s$, that means they must have $K_s - [24] \gtrsim 7-8$. From
Figure \ref{fig:corek24}, the boundary between Flat/Class I is at $K_s - [24] =
8.31$. Therefore, we expect any new YSOs existing among these 634 sources to be
fainter than 18-19 magnitudes at $K_s$ and to be very young Flat/Class I
objects. 

To put an upper limit on the luminosity of any embedded YSOs, we integrated our
median SED assuming a blackbody extrapolation shortward of  $J$ band and
longward of $24\:\mu$m.  The temperature of the blackbody was derived from the
flux ratio of the first two and last two wavelengths, for the shortward and
longward extrapolations, respectively. The luminosity ranges from
$7\times10^{-5}$ L$_\odot$ for L204C-2 (the closest core) to $5\times10^{-4}$
L$_\odot$ for the two furthest cores (L1152 and L1155C-2).

Are any of these sources YSOs?  To answer this question is beyond the scope of
this paper.  However, given that the sources are distributed throughout the
observed regions, it is very likely that most, if not all, are background
galaxies.  Spectra of individual sources would be needed to identify any YSOs
hidden among the  galaxies.

\section{\label{sec:dust} Dust Properties}

The dust properties affect the extinction law.  So, to study the dust
properties, we will compute the extinction law.  By using the extinctions
computed from the \jhks{} bands we can extrapolate the extinction law in the
\spitzer{} wavebands. Starting from Equation \ref{eq:fit}, but using \ak{}
instead of $A_V$, we re-arrange it to solve for $C_{ext}$, defined as
$A_{\lambda}/A_{K_s}$:

\begin{equation}
C_{ext}(\lambda) = \frac{2.5}{A_{K_s}}[\log(k) -
\log(F_{obs}(\lambda)/F_{model}(\lambda))]
\end{equation}

Since we are only interested in \emph{differences} in the extinction law, we can
subtract $C_{ext}(K_s)$ to eliminate the need for $k$, the scaling factor.
Furthermore, because $C_{ext}(K_s) \equiv 1$ and $F_{model}(K_s) \equiv 1$ (all
stellar models are scaled relative to $K_s$), the equation simplifies to:

\begin{equation}
\label{eq:cext}
C_{ext}(\lambda) = \frac{2.5}{A_{K_s}}\left[ \log(F_{obs}(K_s)/F_{obs}(\lambda))
   + \log F_{model}(\lambda)\right] + 1
\end{equation}

For $F_{model}$ we will use the average stellar model derived in Appendix
\ref{sec:avgmodel}.

\subsection{ \label{sec:spatiallaw} Spatial Dependence of the Extinction Law }

In this section we will create a two-dimensional map of how the extinction law
changes in each core.  We start by defining a reduced \chisq{} for each star:

\begin{equation}
\label{eq:chi2}
\chi^2 = \frac{1}{n-1}\sum_\lambda^n \left( \frac{C_{ext}^{obs}(\lambda) -
C_{ext}^{model}(\lambda)}{\sigma_\lambda}\right)^2
\end{equation}

\noindent where $C_{ext}^{obs}(\lambda)$ is computed from Equation
\ref{eq:cext}, $C_{ext}^{model}$ is the extinction law for a given dust model,
and $\sigma_\lambda$ is the uncertainty in $C_{ext}^{obs}$. We sum only over the
IRAC bands because, as we will see, the extinction law at $24\:\mu$m does not
fit any current dust models and therefore we excluded this wavelength to avoid
biasing our \chisq{}. Furthermore, we excluded any negative values for \alak{}
as unphysical. We will discuss the biases introduced by this assumption in
\S\,\ref{sec:error}.  We will refer to this reduced \chisq{} as ``\chisq'' in 
this paper.

\begin{figure}
\epsscale{1}
\plotone{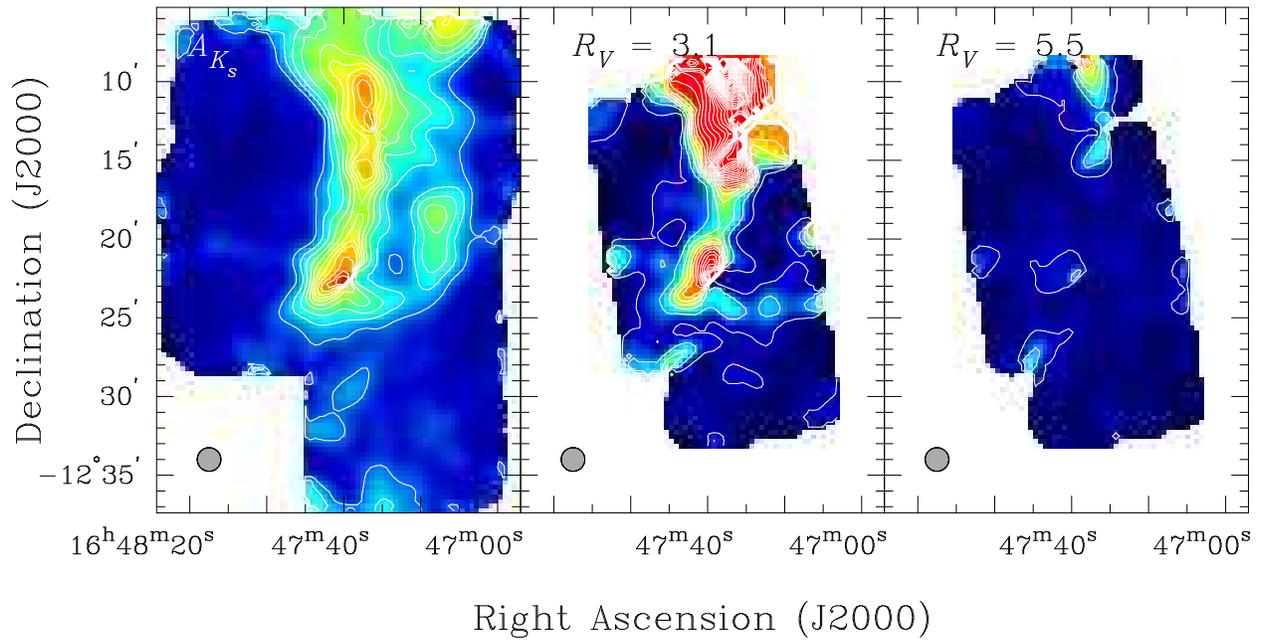}

\caption{\label{fig:l204c-2-chi2} Map of the \chisq{} values in L204C-2. The
left panel shows the \ak{} map with contours starting at 0.5 magnitudes in steps
of 0.15 mag. ($3\sigma$). The middle and right panels show \chisq{} using either
the Weingartner \& Draine $R_V = 3.1$ (middle) or $R_V = 5.5$ (right) dust
models with contours starting at 4 in steps of 4. The gray circle in the lower
left corner denotes the resolution of the maps, $90\arcsec$. }

\end{figure}

\begin{figure}

\plotone{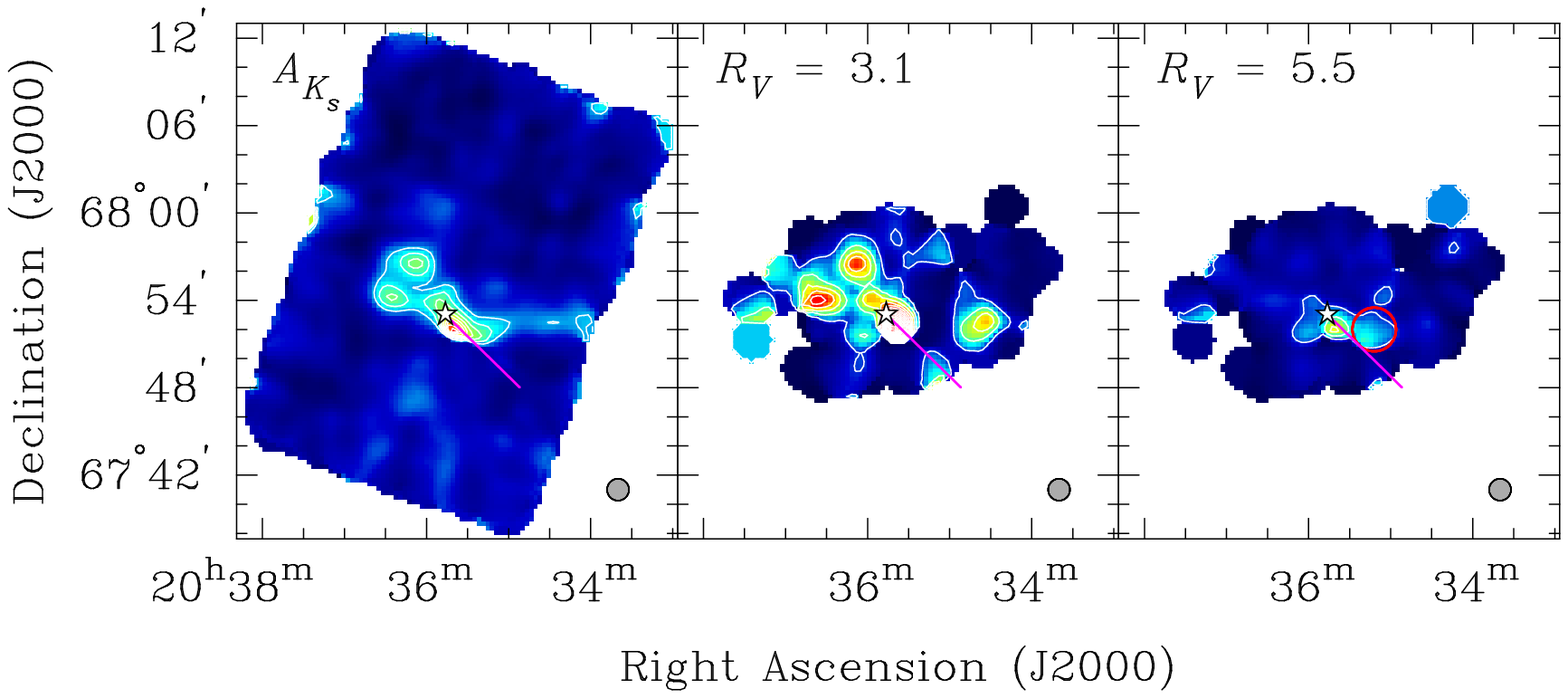}

\caption{\label{fig:l1152-chi2} Same as Figure \ref{fig:l204c-2-chi2} except for
L1152. The star denotes the position of YSO \#1 and the magenta line shows the
direction of the outflow axis.  See \S\,\ref{sec:outflows} for a discussion of
the red circle in the $R_V = 5.5$ panel.}

\end{figure}

\begin{figure}

\plotone{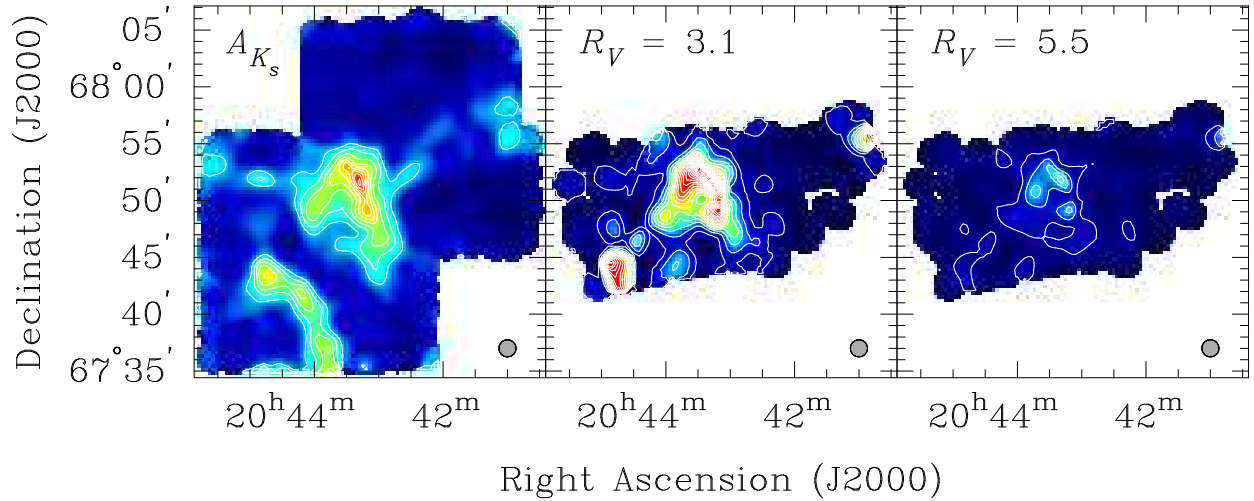}

\caption{\label{fig:l1155c-2-chi2} Same as Figure \ref{fig:l204c-2-chi2} except
for L1155C-2.}

\end{figure}

\begin{figure}

\plotone{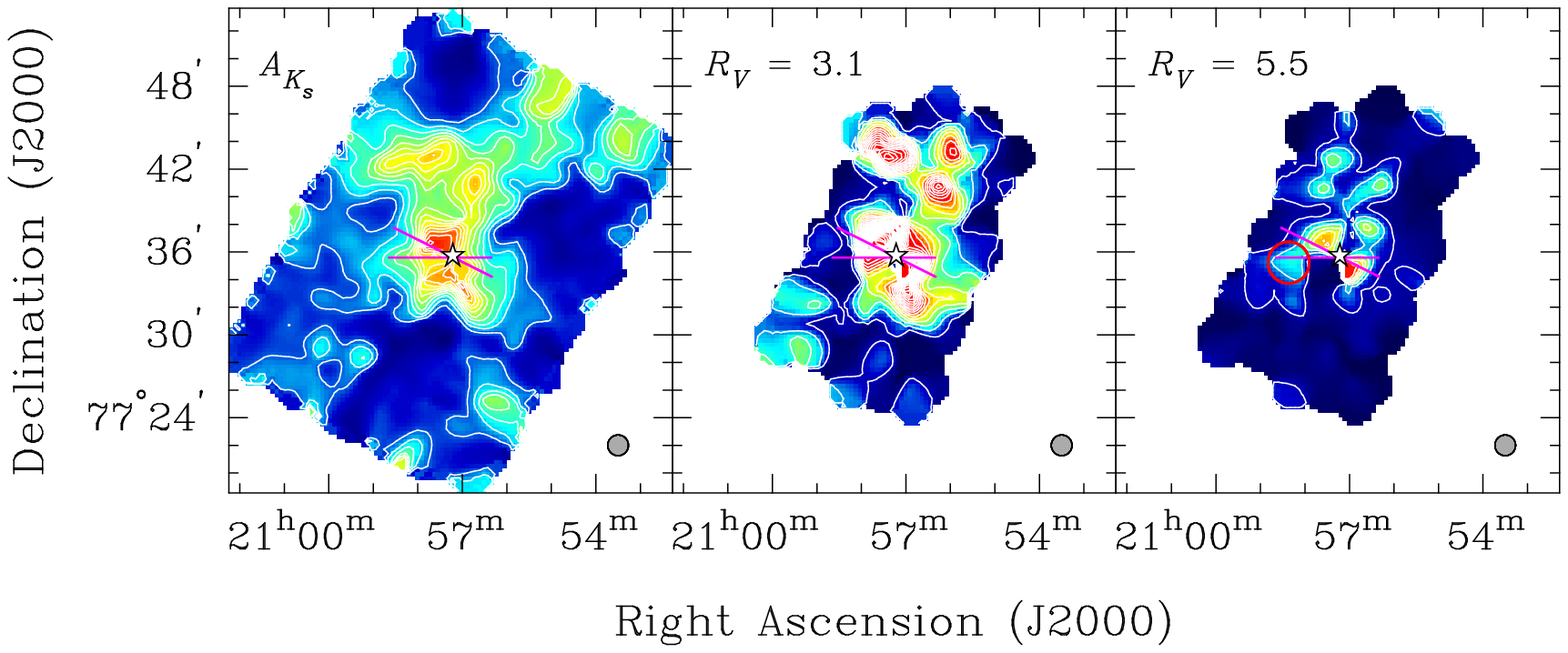}

\caption{\label{fig:l1228-chi2} Same as Figure \ref{fig:l204c-2-chi2} except for
L1228.  The star denotes the position of YSO \#7 and the magenta lines show the
directions of the outflow axes.  See \S\,\ref{sec:outflows} for a discussion of
the red circle in the $R_V = 5.5$ panel.}

\end{figure}

\begin{figure}

\plotone{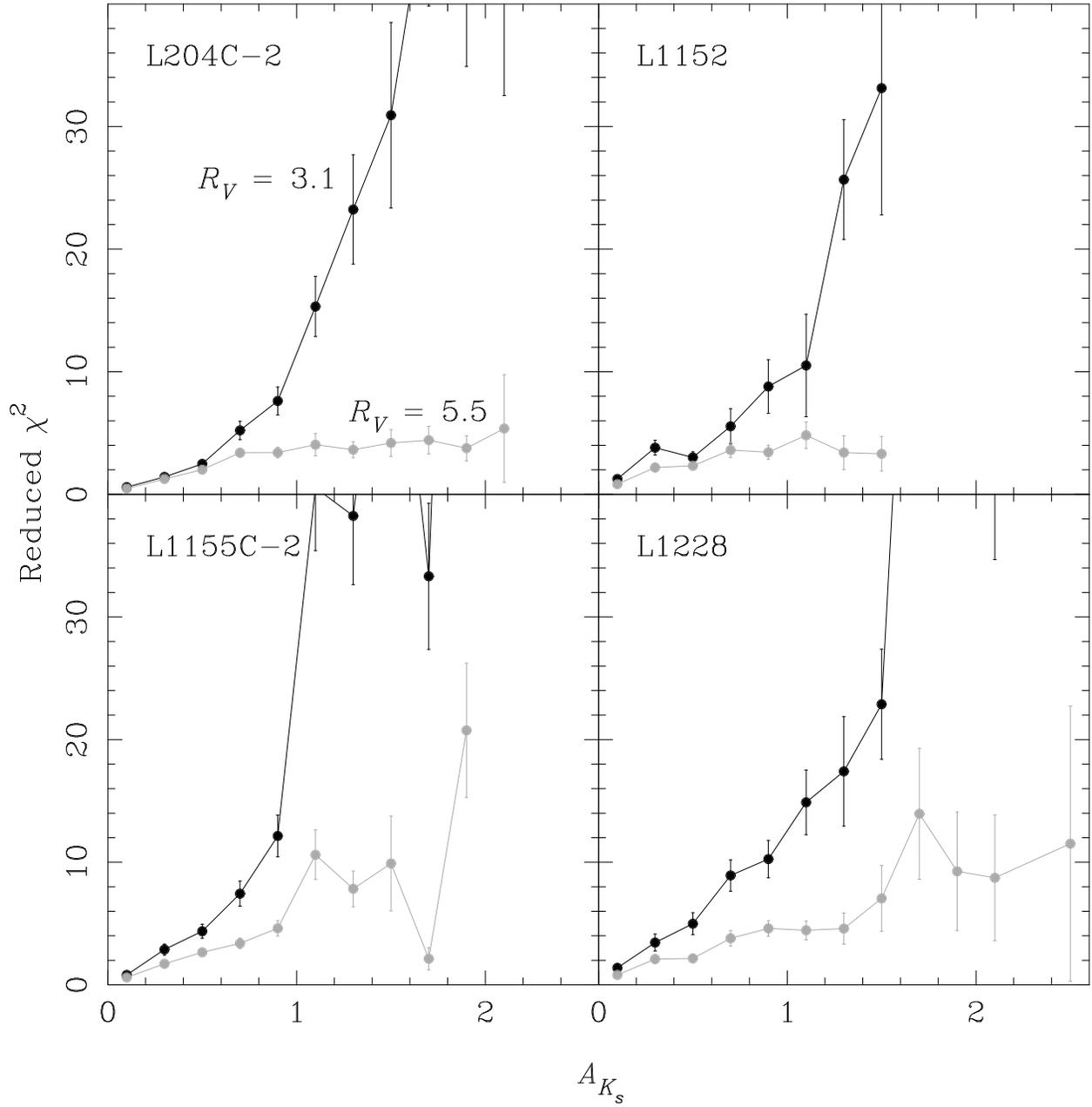}

\caption{\label{fig:cores-chi2ak} \chisq{} versus \ak{} for our cores. The WD3.1
\chisq{} values are black; the WD5.5 \chisq{} values are light gray. The
error bars shown are the standard deviations of the mean in each
bin. }

\end{figure}

Using Equation \ref{eq:chi2} we computed the line-of-sight \chisq{} value for
each star. Then, to convert our line-of-sight measurements into a map, we
followed the same procedure as we did when creating our \ak{} maps. In Figures
\ref{fig:l204c-2-chi2} - \ref{fig:l1228-chi2} we show maps of the extinction and
\chisq{} calculated using both the WD3.1 and WD5.5 dust models. The \chisq{}
maps have the same resolution as the extinction maps, $90\arcsec$. Our contours
start at $\chi^2 = 4$ because we observe a definite transition between the WD3.1
and WD5.5 dust models at this approximate \chisq{} value (see Figure
\ref{fig:cores-chi2ak}).  Statistically, a $\chi^2 = 4$ would arise by chance
about $5\%$ of the time.

The correspondence between the extinction, \ak, and the $R_V = 3.1$ \chisq{}
maps is quite remarkable. In all of our cores, the overall shape and the
extinction peaks are mirrored in the WD3.1 \chisq{} map. Many of the extinction
features not associated with the cores also appear in the \chisq{} map such as
the dusty filament to the north of L204C-2, the two separate extinction peaks in
L1152, and much of the structure in L1228.

To make a quantitative comparison between \chisq{} and \ak{}, we binned our data
in \ak{} and for each bin determined the average \chisq{} with both the WD3.1
and WD5.5 models. Our results are shown in Figure \ref{fig:cores-chi2ak}. $R_V =
3.1$ \chisq{} is shown in black while $R_V = 5.5$ is shown in gray. At low
extinction, $A_{K_s} \lesssim 0.6$, the $R_V = 3.1$ and $R_V = 5.5$ \chisq{}
are very similar to each other, suggesting that at low extinctions it is
difficult to distinguish between these two models with our technique. Above
$A_{K_s} \sim 0.6$ the $R_V = 3.1$ \chisq{} rises sharply while the $R_V = 5.5$
\chisq{} stays roughly constant or increases only slightly.

The observed behavior in our \chisq{} maps is consistent with the idea of grain
growth in dense cores. Most regions of moderate to high extinction show up in
the $R_V = 3.1$ \chisq{} map but not in the $R_V = 5.5$ \chisq{} map. This
suggests the WD3.1 extinction law is \emph{not} valid in extincted regions
because $\chi^2 = 4$ (the first contour level) corresponds to at least a 95\%
chance that the data do not fit the model. The differences in these two
extinction laws is reflected by the differences in the properties of the dust
models. The WD3.1 dust model was constructed to match observations of the
diffuse ISM.  In contrast, the WD5.5 model has significantly fewer small
silicate grains ($r < 0.1\:\mu$m) and significantly more large carbonaceous ones
(maximum radius $\sim 10 \times$ larger).

Although most of our observations can be explained by the idea of grain growth
within dense regions, both L1152 and L1228 contain regions that appear strong in
the WD5.5 \chisq{} map. Both of these cores have outflows associated with them
which may be changing the dust sizes and compositions. In the next section, we
explore changes in the observed extinction law as a function of wavelength and
\ak{}. This will help us to understand exactly how the dust is changing within
our cores.

\subsection{ Wavelength Dependence of the Extinction Law }

As we saw in the last section, the \chisq{} is strongly dependent on \ak. To
explore this further, we started by binning the observed extinction law for each
star into three \ak{} ranges: $0 < A_{K_s} \le 0.5$, $0.5 < A_{K_s} \le 1$, and
$A_{K_s} > 1$. In each extinction bin, we combined the individual
$C_{ext}^{obs}(\lambda)$ measurements to obtain a weighted average value of
\alak. As we did when computing \chisq{}, we again excluded any negative values
of $C_{ext}^{obs}(\lambda)$.

\begin{figure}

\plotone{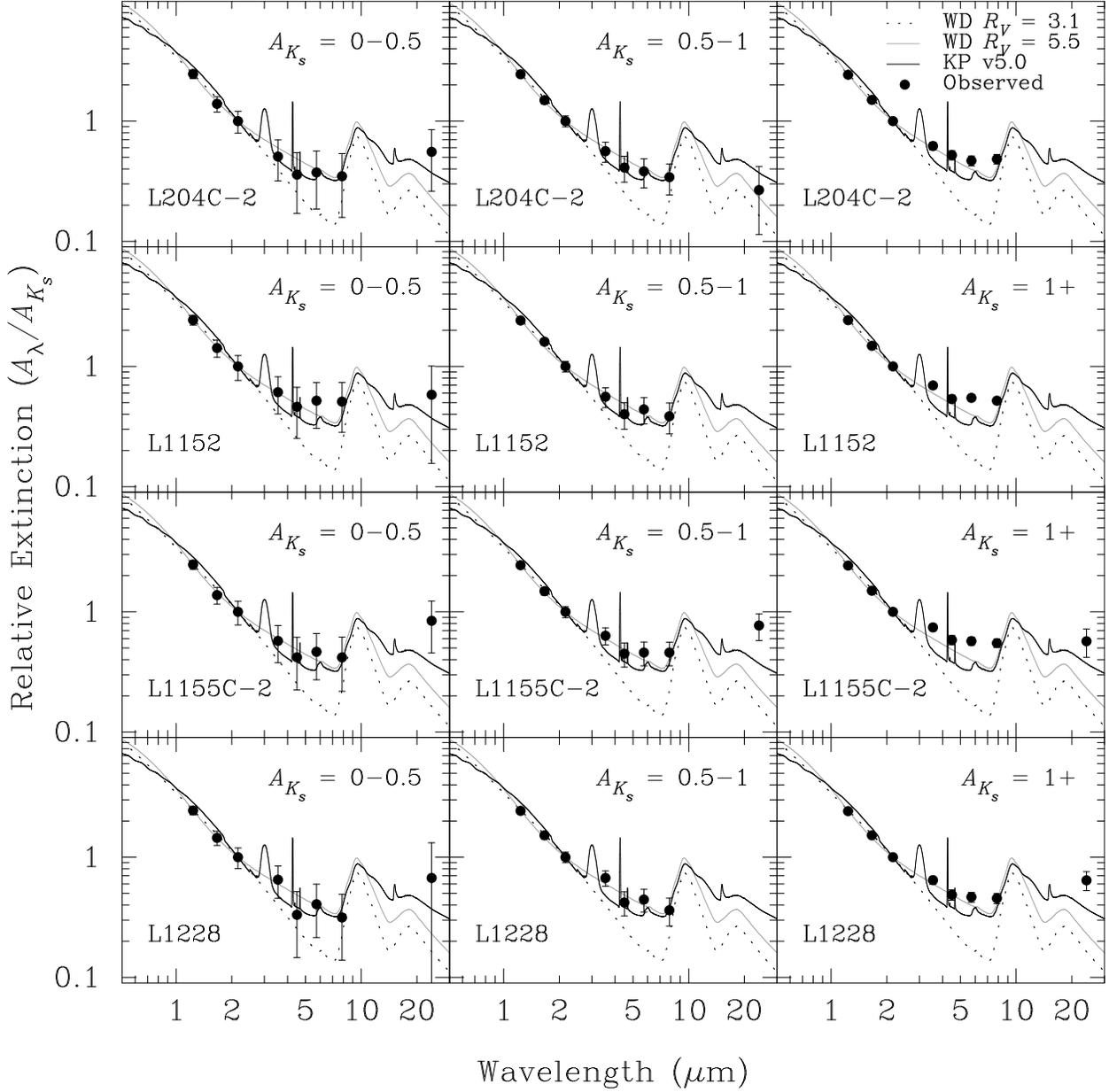}

\caption{\label{fig:coreaklaw} The extinction law for each core divided into
three \ak{} ranges. Extinctions were computed using the WD5.5 dust model with
$\beta = 1.6$ in the \jhks{} bands.  The left column is sources with $0 <
A_{K_s} \le 0.5$, the middle column is $0.5 < A_{K_s} \le 1$, and the right
column is $A_{K_s} > 1$. The errorbars represent the minimum uncertainty in
measuring the fluxes. }

\end{figure}

Figure \ref{fig:coreaklaw} shows the observed average extinction law as a
function of wavelength for our cores. Each row is a different core and each
column is a different range of \ak. The errorbars for each data point are the
minimum uncertainty due to systematic errors in measuring flux.  In addition to
the WD3.1 and WD5.5 dust models, we also plot a third one, labeled KP v5.0
(Pontoppidian et al., in prep). This model is one from a grid of models
constructed starting from the \citet{weingartner01} parameterization of the
grain size distribution. Icy mantles of water and other volatiles were then
added. The specific model we use from this grid is the one with the ``best fit''
to the c2d mid-infrared extinction law and ice features. Several ice absorption
features in this model can be seen in the figures, these are due to H$_2$O,
CO$_2$, or CO.

\subsubsection{\label{sec:spatiallawirac}The Extinction Law From $3.6$ to $8\:\mu$m}

From $3.6$ to $8\:\mu$m, our extinction law is relatively flat for all ranges of
$A_{K_s}$ with the trend that it becomes slightly flatter as $A_{K_s}$
increases. In Table \ref{tab:extlaw} we list the average extinction law combined
over all four cores ($\beta = 1.6$). Again, the uncertainties represent the
minimum error due to uncertainty in the flux.  For comparison we also list the
extinction law from \citet{indebetouw05}, \citet{flaherty07}, \citet{lutz99},
and our three dust models. Our data are in good agreement with other authors'
results and also with the WD5.5 dust model.

Our flat extinction law for $0 < A_{K_s} \le 0.5$ seems surprising since this
result contradicts our expectation that low extinction regions should follow
WD3.1, a diffuse interstellar medium dust model. In \paperone{} we calculated
the extinction law in three molecular clouds and found that for $0 < A_{K_s} \le
0.5$ the extinction law was consistent with WD3.1. These conflicting results may
be because extinction measures the column density, not the number density, along
the line-of-sight.  If we assume each core has an approximate angular size of
$\sim15\arcmin$ (the lengthwise distance of any IRAC observation), and given the
distances listed in Table \ref{tab:corebasic}, then our cores have linear sizes
of $1-3\times 10^5$ AU.  Compare these sizes to those of the clouds in
\paperone{}.  The cloud distances in Table 1 of that paper are $125\pm25$ pc
(Ophiuchus), $250\pm50$ pc (Perseus), and $260\pm10$ pc (Serpens).  Assuming
cloud angular sizes of $\sim2^\circ$, then the clouds have linear sizes of
$1-2\times 10^6$ AU.  Finally, if we assume spherical clouds and cores, then
these linear sizes also represent the depth of each cloud and core. Therefore,
the number densities in our cores may be up to $10\times$ larger than in our
clouds for a given column density (extinction).  Grain growth is a collisional
process and proceeds more rapidly at higher number densities. Hence, the grains
in our isolated cores quickly achieve at least modest grain growth (represented
by WD5.5 and KP v5.0) compared with the WD3.1 dust grains.

L1152 and L1155C-2 both show a strong deviation at $5.8\:\mu$m from WD5.5 for
$A_{K_s} > 1$. Water ice has an absorption peak at $6.02\:\mu$m due to H-O-H
bending \citep{gibb04}. It is possible that the high value of $A_{5.8}/A_{K_s}$
is due to the presence of water ice in these two cores, even more than predicted
by the KP5.0 dust model.  Mid-infrared spectra will be necessary to address this
possibility.

\subsubsection{The Extinction Law at $24\:\mu$m}

We have only 13 stars with positive $A_{24}/A_{K_s}$ values. These have an
average value of $A_{24}/A_{K_s} = 0.59\pm0.12$. This value compares favorably
with other empirical results. \citet{flaherty07} were able to measure the
$24\:\mu$m extinction law for two of their five regions and found
$A_{24}/A_{K_s} = 0.44\pm0.02$ and $0.52\pm0.03$ for Serpens and NGC 2068/71,
respectively. Even though \citet{lutz99} did not measure the extinction longward
of $19\:\mu$m, if their flat extinction law were projected out to $24\:\mu$m, it
would also have a value of $\sim0.5$.  Our result also agrees with \paperone{}
(when a model flux of 0.012 mJy at $24\:\mu$m was assumed).

Although the empirical results agree with each other, they are much higher than
the $24\:\mu$m extinction law predicted by dust models. Our average value for
$A_{24}/A_{K_s}$ is $2-3\times$ larger than either the WD3.1 or WD5.5 models.
The KP v5.0 dust model has a much broader $18\:\mu$m silicate peak which
significantly raises the predicted $24\:\mu$m extinction. Even so, it is still
too low by about $50\%$ compared to our empirical measurement. New dust models
will need to incorporate additional extinction at $24\:\mu$m.

\subsubsection{\label{sec:outflows}Outflows in L1152 and L1228}

\begin{figure}

\plotone{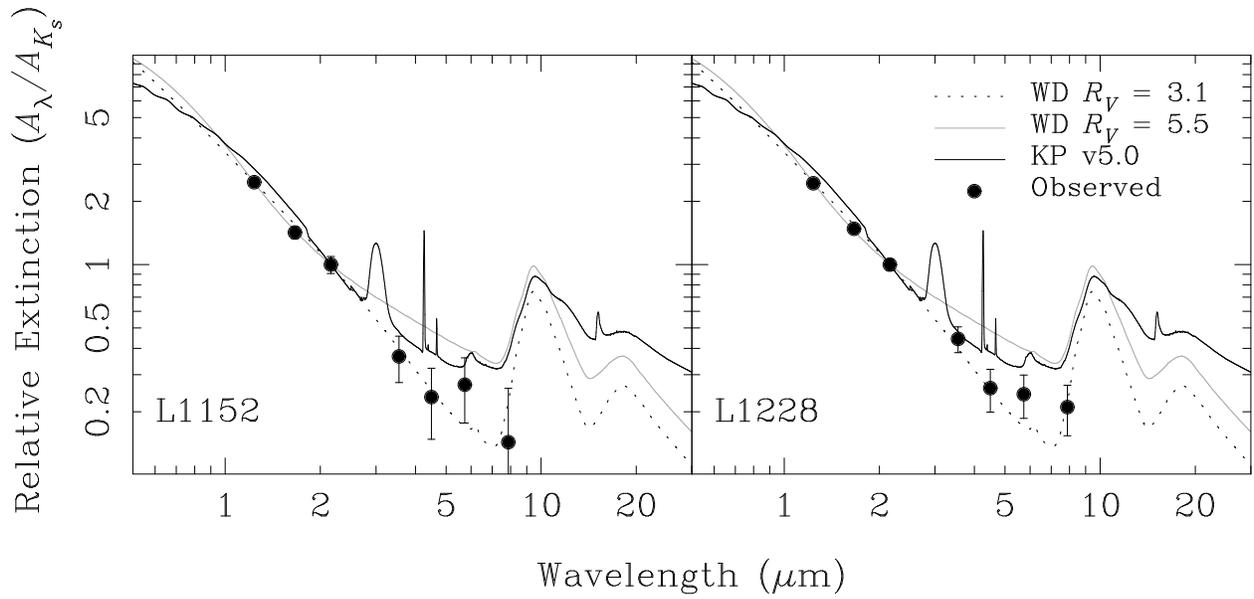}

\caption{\label{fig:aklaw-blobs} The averaged extinction laws for the sources
near the outflows in L1152 (left) and L1228 (right).  Three different dust
models are plotted for comparison.  The errorbars are the minimum uncertainty
due to the systematic errors in the flux. }

\end{figure}

We circled in red two regions in L1152 and L1228 (Figs. \ref{fig:l1152-chi2} and
\ref{fig:l1228-chi2}) with high $R_V=5.5$ \chisq. These regions stand out
because their high \chisq{} values are not caused by a single star, nor are they
associated with regions of high \ak. We selected all the sources within the red
circles and within the $\chi^2 = 4$ contour and plotted the average extinction
law for each region in Figure \ref{fig:aklaw-blobs}.  Compared to elsewhere in
the cores, the extinction law is much steeper in these two regions and more
similar to the WD3.1 dust model rather than WD5.5.  Because these regions are
near known outflows, one possible explanation is that the outflows in each core
have destroyed the bigger dust grains via shocks to produce a dust grain
distribution that more closely resembles the WD3.1 diffuse ISM dust. Since
neither region lines up with the outflow axis, this may be evidence for
precession of the outflows, which is already observed in L1228 \citep{bally95}. 
These extinction laws could not be caused by a foreground star skewing the
average because we have a total of 35 stars in the two selected regions, none
with $A_{K_s} < 0.20$.  A  foreground star would appear to have almost zero or
negative extinction, and would not greatly impact the average extinction law.

\subsection{\label{sec:error} Sources of Error}

\subsubsection{Negative \cext}

In this paper we have excluded negative values for \cext{} as unphysical because
it would imply that dust amplification, rather than extinction, is occurring. A
more likely explanation is that the derived extinction or the measured flux for
that wavelength is incorrect.  By excluding these negative values, we introduce
some potential bias into our results, which will be discussed here.

The magnitude of the bias varies with wavelength and extinction, but is much
larger at low \ak{}.  The percentage of negative \cext{} values for a given
wavelength ranges from $32-61\%$ for $0 < A_{K_s} \leq 0.5$, $3-17\%$ for $0.5 <
A_{K_s} \leq 1$, and $0-1.8\%$ for $A_{K_s} > 1$.  If we utilized these negative
\cext{} values when computing the average extinction law, then the averages shown
in Figure \ref{fig:coreaklaw} would be $0.42-1.08\sigma$ lower for $0 < A_{K_s}
\leq 0.5$, where $\sigma$ is  the displayed errorbar due to the minimum
uncertainty in the flux.  For  higher extinctions, the bias is much lower because
there are far fewer  negative \cext{} values.  The average extinction law
decreases $0.05-0.52\sigma$ for $0.5 < A_{K_s} \leq 1$ and for $A_K{_s} > 1$, the
decrease is $0-0.06\sigma$.

The large bias we find for $0 < A_{K_s} \leq 0.5$ suggests that our technique
for computing the extinction law breaks down at low extinctions.  We discussed
this before in relation to Figure \ref{fig:cores-chi2ak}.  We have not attempted
to incorporate any bias from excluding the negative \cext{} into our errorbars
since the bias is partially due to the fluxes, and we already set a lower limit
on the errors due to our minimum uncertainty in measuring the fluxes at
different wavelengths.  However, our results presented in this paper may
slightly overestimate the mid-infrared extinction law, primarily for $A_{K_s}
\leq 0.5$.

\subsubsection{ \label{sec:beta} Changes in the Near-Infrared Extinction Law}

\begin{figure}

\plotone{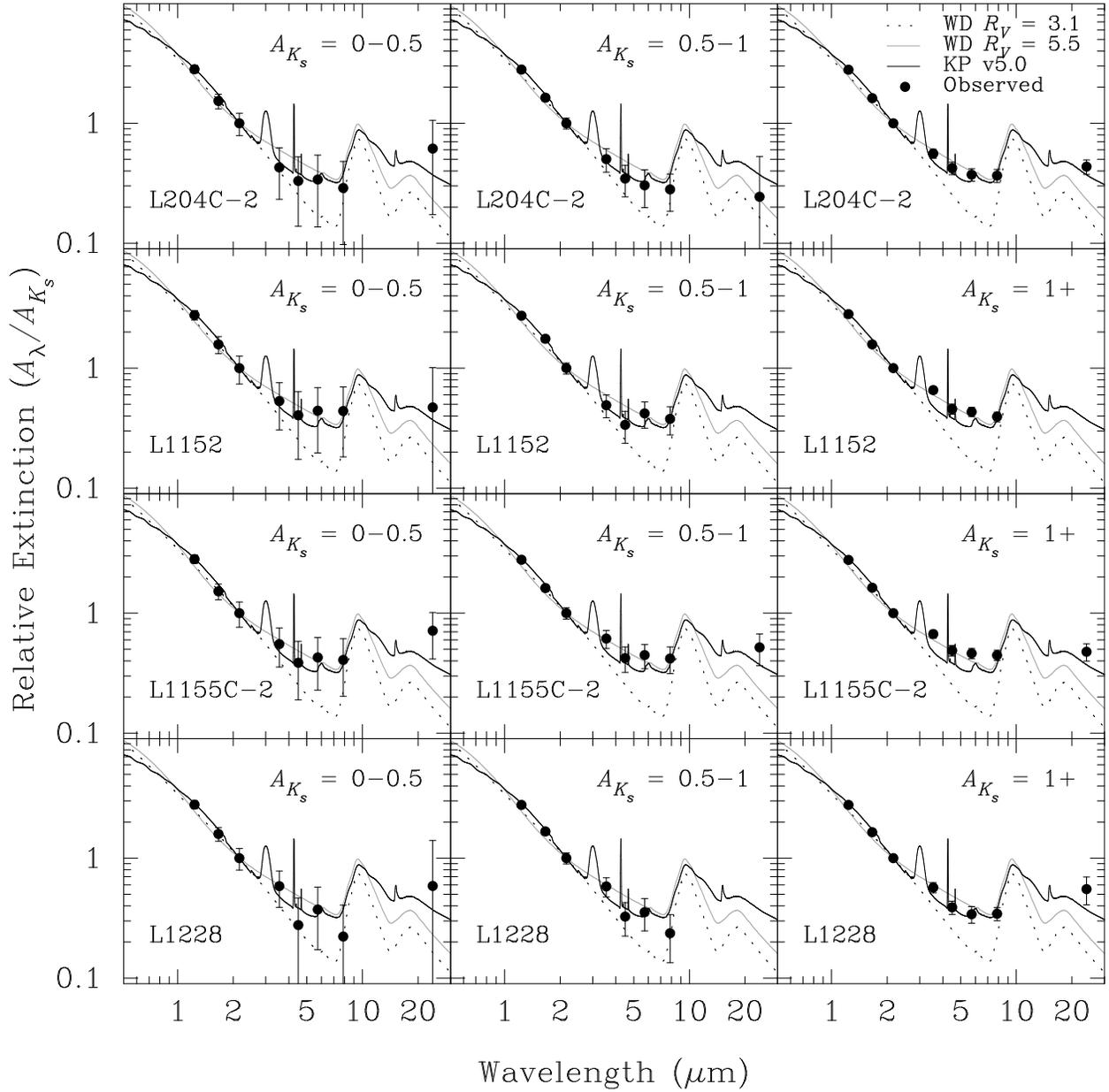}

\caption{\label{fig:beta18} Same as Figure \ref{fig:coreaklaw} except using an
extinction law with $\beta = 1.8$ in the \jhks{} bands for computing extinctions
and source classification.}

\end{figure}

Many authors have fit the near-infrared extinction to a power law, $A_\lambda
\propto \lambda^{-\beta}$, with $\beta = 1.6-1.8$ \citep[and references
therein]{draine03}.  The WD3.1 and WD5.5 models are very similar in the
near-infrared \jhks{} bands, reflecting this apparent universality of the
near-infrared extinction law. These two models both have $\beta \approx 1.6$.
This law does agree with the  apparent reddening vector in our data (see Figure
\ref{fig:ahak-withgal}) and makes our results directly comparable with other
authors' results who have made the same assumption.  However, it is possible
that the extinction law could be steeper in some regions.  This would introduce
another source of error into our calculated extinction law.  To calculate the
magnitude of this effect, we modified the WD5.5 dust model to have $\beta = 1.8$
in the \jhks{} bands and used this law instead of the WD5.5 model for
classifying sources and computing the extinctions and the extinction law.  Our
results can be seen in Table \ref{tab:extlaw} and Figure \ref{fig:beta18}.  As
stated before, the uncertainties in our average extinction law are the minimum
errors due to uncertainty in the flux.

Using a $\beta = 1.8$ law slightly decreases our mid-infrared extinction law,
however the qualitative results of this paper are unchanged.  Even with
$\beta=1.8$ our extinction law is more consistent with WD5.5 for all ranges of
extinction.  The biggest difference is for $A_{K_s} > 1$ in L1152 and L1155C-2.
With $\beta=1.8$, the excess $A_{5.8}/A_{K_s}$ extinction disappears. We argued
in \S\,\ref{sec:spatiallawirac} that this excess could possibly be explained by
the existence of water ice in these two cores.  The circled regions in our
\chisq{} figures that may be caused by outflows destroying the big dust grains
appear even stronger with our $\beta = 1.8$ law.  This is because the 
difference from WD5.5 is enhanced.

\section{\label{sec:conclude}Conclusions}

In this paper we presented deep \jhks{} and \spitzer{} photometry of four
isolated cores: L204C-2, L1152, L1155C-2, and L1228. Based on previous
observations, two of these cores were classified as starred and two were
classified as starless. Our data support these original classifications. We
detect 10 YSOs, three within L1152, seven within L1228, and none within L204C-2
or L1155C-2. Seven of these YSOS have not been previously discovered. Among
these, we are able to resolve the driving source of the HH200 outflow in L1228
into two sources separated by $5\farcs3$, but we were unable to identify which
of the two sources is responsible for the outflow.

To put constraints on possible faint YSOs in all four cores, we selected 634
sources with $24\:\mu$m detections $\ge 3\sigma$ which are either undetected or
too faint at other wavelengths to be considered in our high reliability
catalogs.  We computed the median SED for these sources and integrated the  SED
to obtain a luminosity of $7\times 10^{-5}$ to $5\times 10^{-4}$ L$_\odot$ with
colors of a very young Flat spectrum or Class I YSO \citep{greene94}.

In addition to studying the YSO content of these cores, we used the
line-of-sight measurements of the extinction and the extinction law to create
maps of the extinction and \chisq{} deviation from specific dust grain models.
Our extinction law is nearly flat for all ranges of extinction in all four
cores.  In the IRAC bands ($3.6-8\:\mu$m) this extinction law matches the
predictions of the WD5.5 dust model, a model designed to simulate the grain
growth that occurs in dense regions, and also agrees with the other authors'
results \citep{indebetouw05,flaherty07,lutz99}.  In the densest regions of L1152
and L1155C-2, the $5.8\:\mu$m extinction law is higher than the WD5.5 model.
This could be evidence for water ice in these cores or a steeper near-infrared
extinction law than the average value.  At $24\:\mu$m, our data are consistent
with the extinction law remaining constant as the total extinction increases.
The observed extinction law is much higher than that predicted by dust grain
models, but does confirm the similarly large value found by other authors
\citep{chapman09,flaherty07,lutz99}.

From the \chisq{} maps we identified cavities in L1152 and L1228 where the
extinction law is more consistent with the WD3.1 dust model.  The molecular
outflows in these cores produce shocks, which may be destroying the large dust
grains, thus producing a dust distribution similar to WD3.1 These cavities are
visible in our \chisq{} maps, but would be missed in averages over an entire
core.

It is critical to create new dust grain models which reproduce the observed flat
extinction law, especially at $24\:\mu$m.  In this paper we used the
\citet{robitaille07} models to estimate the luminosities of our YSOs.  However,
\citet{robitaille07} assumed an extinction law that is between WD3.1 and WD5.5
in the IRAC bands and predicts a $24\:\mu$m extinction that is about $25\%$ of
our observed value.  The best-fit YSO parameters would undoubtedly change with
a  flatter extinction law.  Furthermore, it is important to understand the
mid-infrared extinction law because future telescopes, such as the James Webb
Space Telescope, will operate at these wavelengths.

\acknowledgements

We are grateful to Neal Evans for providing useful feedback on several drafts
and also the anonymous referee whose comments greatly improved the quality of
this paper.  Support for this work, utilizing data from the ``Cores to Disks''
Spitzer Legacy Science Program \citep{evans03}, was provided by NASA through
contracts 1224608, 1230782, and 1230779 issued by the Jet Propulsion Laboratory,
California Institute of Technology, under NASA contract 1407. Additional support
for N.L.C.\ was provided by NASA through JPL contracts 1264793 and 1264492.
L.G.M.\ was supported by NASA Origins Grant NAG510611. This research has made
use of PyRAF, a product of the Space Telescope Science Institute, which is
operated by AURA for NASA. We also used the SIMBAD  database which is operated
at CDS, Strasbourg, France.

\appendix
\section{ \label{sec:avgmodel} Average Stellar Models }

To compute the expected stellar distribution for each core, we used the galaxy
model from \citet{jarrett92} and \citet{jarrett94}. Given the Galactic
coordinates, distance, and average extinction for a core, this model produces
source counts broken down by spectral type. These source counts provide a
rudimentary weighting function that can be used in computing the average stellar
model. However, a weighting function from the raw source counts ignores the
reality that not all spectral types are equally observable at each wavelength.
Our next step was to correct for the detection limits of our actual
observations. The galaxy model lists the extincted $K_s$ magnitude for each star
and its extinction, $A_V$. We used this information, the \citet{weingartner01}
$R_V = 3.1$ extinction law, and the stellar templates from the SSC's
``star-pet''\footnote{\url{http://ssc.spitzer.caltech.edu/tools/starpet}} tool
to compute the magnitude of each star at every observed waveband from $J$ to
$24\:\mu$m.  We then arbitrarily discarded any detection falling below the
$5\sigma$ cutoff for the given waveband (Table \ref{tab:limits}).  Lastly, we
used the normalized, final, corrected source counts at each waveband as a
weighting function, and computed the overall average stellar model.  We repeated
this procedure using both the \citet{weingartner01} $R_V = 5.5$ extinction law
and also using the magnitude limits from our deeper \jhks{} observations.  Both
variables have a negligible effect on the final fluxes.

After these calculations, we made one final adjustment to our average stellar
model. In \paperone{} we determined that the above method produces a $24\:\mu$m
flux that is very likely too high. However, we can independently put an
upper-limit on the $24\:\mu$m flux value. If we select all stars with
line-of-sight $0 < A_{K_s} \le 0.5$ and compute the average $24\:\mu$m to $K_s$
flux ratio, this will give us an estimate of the true average $24\:\mu$m stellar
flux. Since these stars have small amounts of extinction, this ratio will
over-estimate the true value because the $K_s$ band is more extincted than
$24\:\mu$m. There are 28 stars in our cores with $24\:\mu$m fluxes $\geq
3\sigma$ and within the extinction range $0 < A_{K_s} \le 0.5$. These have an
average $24\:\mu$m to $K_s$ flux of $0.013\pm0.001$ with an average extinction
of 0.19 $A_{K_s}$. This is the same as the value derived using the Galaxy count
models. Given the non-zero average extinction, the true flux ratio will be
lower. Therefore, in this paper we will assume it is actually 0.012. Our final
average stellar flux for the \jhks, IRAC1-4, and MIPS1 bands, respectively, is:
$1.282\pm0.144$, $1.302\pm0.050$, $1.$, $0.453\pm0.010$, $0.268\pm0.005$,
$0.179\pm0.005$, $0.102\pm0.003$, and $0.012\pm0.001$. Note that all fluxes are
scaled relative to the $K_s$ band.

\begin{deluxetable}{lccc}
\tablewidth{0pt}
\tablecolumns{4}
\tablecaption{\label{tab:corebasic} Basic Properties of the Cores }
\tablehead{\colhead{} & \colhead{$l$} & \colhead{$b$} & \colhead{Dist.} \\
\colhead{ Core } & \colhead{(deg.)} & \colhead{(deg.)} & \colhead{(pc)}}
\startdata
L204C-2  &   6 & 20 & $125\pm25$\tablenotemark{a}\\
L1152    & 102 & 16 & $325\pm25$\tablenotemark{b}\\
L1155C-2 & 102 & 15 & $325\pm25$\tablenotemark{b}\\
L1228    & 111 & 20 & $200\pm50$\tablenotemark{c}
\enddata
\tablenotetext{a}{\citet{degeus89}}
\tablenotetext{b}{\citet{straizys92}}
\tablenotetext{c}{\citet{kun98}}
\end{deluxetable}

\begin{deluxetable}{lccc}
\tablewidth{0pt}
\tablecolumns{4}

\tablecaption{\label{tab:coreobs} Summary of \spitzer{} Observations }

\tablehead{\colhead{ Core } & \colhead{ AOR Number } & \colhead{ Date Observed }
         & \colhead{ Program ID }\\
\colhead{} & \colhead{} & \colhead{YYYY-MM-DD} & \colhead{}}
\startdata

L204C-2  & 11393792 & 2005-04-07 & 3656\\
         & 11396352 & 2005-04-08 & 3656\\
         & 11392000 & 2005-08-22 & 3656\\
         & 11398912 & 2005-08-23 & 3656\\
L1152    & 11390976 & 2004-07-23 & 3656\\
         & 11399424 & 2004-07-28 & 3656\\
         & 11394304 & 2004-10-15 & 3656\\
         & 11396864 & 2004-11-10 & 3656\\
L1155C-2 & 11392768 & 2004-07-22 & 3656\\
         & 11399936 & 2004-08-12 & 3656\\
         & 11394816 & 2004-12-02 & 3656\\
         & 11397376 & 2004-12-26 & 3656\\
L1228    & 11391232 & 2004-11-28 & 3656\\
         & 11395072 & 2004-12-07 & 3656\\
         & 11400192 & 2004-12-16 & 3656\\
         & 11397632 & 2004-12-26 & 3656\\

\enddata
\end{deluxetable}

\begin{deluxetable}{cccc}
\tablewidth{0pt}
\tablecolumns{4}
\tablecaption{\label{tab:limits} Sensitivity Limits}
\tablehead{\colhead{Band} & \colhead{Central $\lambda$} &
\colhead{$10\:\sigma$} & \colhead{$5\:\sigma$}\\
\colhead{} & \colhead{($\mu$m)} & \colhead{(mag.)} & \colhead{(mag.)}}
\startdata
$J$   & 1.235 & 19.5(19.9) & 20.3(20.7)\\
$H$   & 1.662 & 18.8(19.7) & 19.5(20.4)\\
$K_s$ & 2.159 & 17.7(18.6) & 18.4(19.3)\\
IRAC1 & 3.550 & 18.6       & 19.7\\
IRAC2 & 4.493 & 17.8       & 18.8\\
IRAC3 & 5.731 & 15.5       & 16.4\\
IRAC4 & 7.872 & 14.7       & 15.6\\
MIPS1 &  23.7 & 10.4       & 11.7\\
\enddata
\tablecomments{The parenthetical values are the limits obtained for our deeper
observations towards the dense core regions L1152 and L1155C-2.}
\end{deluxetable}

\begin{deluxetable}{rlccccccccccc}
\tablewidth{0pt}
\tablecolumns{13}
\rotate
\tabletypesize{\small}

\tablecaption{\label{tab:coreysos} Photometry of Young Stellar Objects }

\tablehead{\colhead{} & \colhead{} & \colhead{ RA } & \colhead{ DEC } &
\colhead{$J$} & \colhead{$H$} & \colhead{$K_s$} & \colhead{$3.6\:\mu$m} &
\colhead{$4.5\:\mu$m} & \colhead{$5.8\:\mu$m} & \colhead{$8.0\:\mu$m} &
\colhead{$24\:\mu$m} & \colhead{$70\:\mu$m}\\
\colhead{Index} & \colhead{Core} & \colhead{(J2000)} & \colhead{(J2000)} &
\colhead{(mJy)} & \colhead{(mJy)} & \colhead{(mJy)} & \colhead{(mJy)} &
\colhead{(mJy)} & \colhead{(mJy)} & \colhead{(mJy)} & \colhead{(mJy)} &
\colhead{(mJy)}}

\startdata

1 & L1152 & 20 35 46.3 & 67 53 02.2& 0.130 & 0.696 & 1.57 & 3.59 & 6.44 &
5.50 & 3.36 & 280 & 3670\\

2 & L1152 & 20 36 11.6 & 67 57 09.3& 9.19  & 25.5  & 36.1 & 41.3 & 47.9 &
60.8 & 77.8 & 183 & 347\\

3 & L1152 & 20 36 19.9 & 67 56 31.6& 52.8  & 130.  & 205  & 228  & 243  &
259  & 262  & 597 & 1080\\

4 & L1228 & 20 55 37.1 & 77 38 19.6& 5.76  & 7.18  & 7.21 & 5.69 & 4.68 &
3.78 & 3.89 & 4.58& \nodata\\

5 & L1228 & 20 57 06.8 & 77 36 56.1& 0.541 & 1.78  & 3.13 & 7.81 & 11.1 &
13.9 & 18.5 & 271 & 2240\\

6 & L1228 & 20 57 08.0 & 77 36 59.7& 0.624 & 1.91  & 2.82 & 3.16 & 3.46 &
3.28 & 7.32 & 154 & \nodata\\

7 & L1228 & 20 57 13.0 & 77 35 43.3& 8.74  & 58.5  & 143  &  262 &  374 &
499  & 762  & 2680& 7850\\

8 & L1228 & 20 57 15.5 & 77 34 23.6& 0.141 & 0.715 & 1.79 & 2.00 & 2.41 &
2.46 & 2.94 & 5.04& \nodata\\

9 & L1228 & 20 57 17.0 & 77 36 58.6& 5.14  & 15.1  & 26.1 & 31.3 & 38.0 &
37.2 & 42.1 & 83.8& 334\\

10& L1228 & 20 58 40.0 & 77 27 45.5& 1.17  & 3.08  & 6.53 & 22.9 & \nodata &
24.7 & \nodata & 43.6 & \nodata
\enddata

\end{deluxetable}

\begin{deluxetable}{rccccc}
\tablewidth{0pt}
\rotate
\tablecolumns{6}
\tablecaption{\label{tab:robitaille} Selected Model YSO Parameters }
\tablehead{\colhead{Source} & \colhead{Envelope Accretion Rate} & 
\colhead{Disk Mass} & \colhead{Interstellar $A_V$} & \colhead{$L_{bol}$} & 
\colhead{YSO} \\
\colhead{ID} & \colhead{($10^{-6}$ $M_\odot$/year)} & \colhead{$10^{-6}$ $M_\odot$} & 
\colhead{(mag)} & \colhead{(L$_\odot$)} & \colhead{Class\tablenotemark{a}}}

\startdata
 1 & $0.05-26.15$ $(10.18)$ & 
     $10 - 11,000$ $(1,600)$ &
     $ 4.7-15.6$ $(10.6)$ & $  0.52-  3.17$ $( 1.19)$ & I\\
 2 & $0 - 0.10$   $(0.05)$  &
     $230 - 5,300$ $(2,800)$ &
     $ 4.0-10.8$ $( 7.4)$ & $  2.06-  2.68$ $( 2.37)$ & II\\
 3 & $0 - 0.16$   $(0.07)$  &
     $590 - 85,000$ $(14,000)$ &
     $ 6.5-10.5$ $( 8.9)$ & $  7.61- 21.25$ $(11.18)$ & II\\
 4 & $0 - 0$      $(0)$     &
     $3.6-18$ $(11)$ &
     $ 1.5- 1.8$ $( 1.7)$ & $  0.03-  0.03$ $( 0.03)$ & II\\
 5 & $0.21-4.38$  $(2.36)$  &
     $14-760$ $(420)$ &
     $ 3.4-12.9$ $( 7.1)$ & $  0.35-  3.21$ $( 0.93)$ & I\\
 6 & $0 - 7.15$   $(0.32)$  &
     $3.4-12,000$ $(880)$ &
     $ 2.8-12.7$ $(10.7)$ & $  0.04-  3.31$ $( 0.20)$ & I\\
 7 & $0 - 0.11$   $(0.07)$  &
     $8,900-26,000$ $(20,000)$ &
     $16.5-16.8$ $(16.6)$ & $  9.63- 12.57$ $(10.61)$ & Flat\\
 8 & $0 - 0$      $(0)$     &
     $0.55-860$ $(89)$ &
     $11.9-16.4$ $(15.0)$ & $  0.03-  0.15$ $( 0.04)$ & II\\
 9 & $0.13-0.13$  $(0.13)$  &
     $15-15$ $(15)$ &
     $ 9.4-10.3$ $( 9.8)$ & $  0.29-  0.29$ $( 0.29)$ & II\\
10 & $0 - 19.07$  $(4.99)$  &
     $8.3-58,000$ $(4,600)$ &
     $ 0.1-12.3$ $( 4.8)$ & $  0.24-  5.19$ $( 0.96)$ & Flat\\
\enddata
\tablenotetext{a}{\citet{greene94}}
\tablecomments{The parenthetical values denote the average value for each
quantity}
\end{deluxetable}

\begin{deluxetable}{lccccc}
\rotate
\tablewidth{0pt}
\tablecolumns{6}
\tablecaption{ \label{tab:extlaw} Average Relative Extinction \alak}
\tablehead{\colhead{Source} & \colhead{$3.6\:\mu$m} & \colhead{$4.5\:\mu$m} &
\colhead{$5.8\:\mu$m} & \colhead{$8\:\mu$m} & \colhead{$24\:\mu$m}}

\startdata
All cores, $\beta = 1.6$ & $0.64\pm0.07$ & $0.49\pm0.06$ & $0.48\pm0.06$ &
$0.47\pm0.05$ & $0.59\pm0.12$ \\
All cores, $\beta = 1.8$ & $0.57\pm0.08$ & $0.40\pm0.07$ & $0.40\pm0.06$ &
$0.37\pm0.05$ & $0.47\pm0.06$\\
\citet{flaherty07}   & $0.632 \pm 0.004$ & $0.54 \pm 0.01$ &
$0.50 \pm 0.02$ & $0.50 \pm 0.01$ & $0.46 \pm 0.04$\\
\citet{indebetouw05} & $0.56 \pm 0.06$   & $0.43 \pm 0.08$ &
$0.43 \pm 0.10$ & $0.43 \pm 0.10$ & \nodata\\
\citet{lutz99}\tablenotemark{a}&$0.53 \pm 0.03$   & $0.50 \pm 0.08$ &
$0.49 \pm 0.06$ & $0.42 \pm 0.06$ & \nodata\\
WD3.1\tablenotemark{b} & 0.40 & 0.25 & 0.17 & 0.22 & 0.17\\
WD3.1\tablenotemark{b} & 0.60 & 0.49 & 0.40 & 0.41 & 0.24\\
KP, v5.0\tablenotemark{b} & 0.48 & 0.38 & 0.34 & 0.38 & 0.38\\
\enddata
\tablenotetext{a}{The extinction for the closest ISO wavelength to each
\spitzer{} band is listed: 3.7, 4.4, 5.9, and $7.5\:\mu$m}
\tablenotetext{b}{Extinctions for dust models are computed at the central
wavelength in each band}

\end{deluxetable}

\clearpage

\clearpage

\end{document}